\def\year{2019}\relax
\documentclass[letterpaper]{article} 
\usepackage{aaai19}  
\usepackage{times}  
\usepackage{helvet} 
\usepackage{courier}  
\usepackage[hyphens]{url}  
\usepackage{graphicx} 
\usepackage{graphicx}  
\usepackage{type1cm}     
\usepackage{graphicx}     
\usepackage{xspace}     
\usepackage{booktabs}     
\usepackage{multirow}      
\usepackage[vlined,linesnumbered,ruled,noend]{algorithm2e}     
\usepackage{microtype}    
\usepackage{units}     
\usepackage{mathtools}     
\usepackage{amsmath}
\usepackage{amssymb}     
\usepackage{xfrac}    
\usepackage[font={small}]{subfig} 
\usepackage{enumitem}
\usepackage{makecell}
\usepackage{mathtools,xparse}

\usepackage{hyphenat}

\usepackage{mathptmx}

\DeclarePairedDelimiter{\norm}{\lVert}{\rVert}

\NewDocumentCommand{\normL}{ s O{} m }{%
    \IfBooleanTF{#1}{\norm*{#3}}{\norm[#2]{#3}}_{F}^2%
}
\NewDocumentCommand{\normF}{ s O{} m }{%
     \IfBooleanTF{#1}{\norm*{#3}}{\norm[#2]{#3}}_{F}%
}
\NewDocumentCommand{\normsquare}{ s O{} m }{%
    \IfBooleanTF{#1}{\norm*{#3}}{\norm[#2]{#3}}^2%
}

\newcommand{\spara}[1]{\smallskip\noindent\textbf{#1.}}

\newcommand{\cpt}[1]{\textsc{\MakeLowercase{#1}}}
\newcommand{\verified}{\cpt{Verified}\xspace}
\newcommand{\political}{\cpt{Political}\xspace}
\newcommand{\despoina}{\cpt{Despoina}\xspace}
\newcommand{\despoinarandom}{\cpt{Despoina (random)}\xspace}

\newcommand{\hide}[1]{}

\newcommand{\eg}{{e.g.}\xspace}
\newcommand{\cf}{{cf.}\xspace}

\newcommand{\vs}{{vs.}\xspace}

\usepackage{pifont}

\newcommand{\xmark}{\ding{55}}

\newcommand{\textcite}[1]{\citeauthor{#1} \shortcite{#1}}

\DeclareSymbolFont{greek}{OML}{cmm}{m}{n}
\DeclareMathSymbol{\alpha}{\mathalpha}{greek}{"0B}
\DeclareMathSymbol{\beta}{\mathalpha}{greek}{"0C}
\DeclareMathSymbol{\gamma}{\mathalpha}{greek}{"0D}
\DeclareMathSymbol{\delta}{\mathalpha}{greek}{"0E}
\DeclareMathSymbol{\epsilon}{\mathalpha}{greek}{"0F}
\DeclareMathSymbol{\zeta}{\mathalpha}{greek}{"10}
\DeclareMathSymbol{\eta}{\mathalpha}{greek}{"11}
\DeclareMathSymbol{\theta}{\mathalpha}{greek}{"12}
\DeclareMathSymbol{\iota}{\mathalpha}{greek}{"13}
\DeclareMathSymbol{\kappa}{\mathalpha}{greek}{"14}
\DeclareMathSymbol{\lambda}{\mathalpha}{greek}{"15}
\DeclareMathSymbol{\mu}{\mathalpha}{greek}{"16}
\DeclareMathSymbol{\nu}{\mathalpha}{greek}{"17}
\DeclareMathSymbol{\xi}{\mathalpha}{greek}{"18}
\DeclareMathSymbol{\pi}{\mathalpha}{greek}{"19}
\DeclareMathSymbol{\rho}{\mathalpha}{greek}{"1A}
\DeclareMathSymbol{\sigma}{\mathalpha}{greek}{"1B}
\DeclareMathSymbol{\tau}{\mathalpha}{greek}{"1C}
\DeclareMathSymbol{\upsilon}{\mathalpha}{greek}{"1D}
\DeclareMathSymbol{\phi}{\mathalpha}{greek}{"1E}
\DeclareMathSymbol{\chi}{\mathalpha}{greek}{"1F}
\DeclareMathSymbol{\psi}{\mathalpha}{greek}{"20}
\DeclareMathSymbol{\omega}{\mathalpha}{greek}{"21}
\DeclareMathSymbol{\varepsilon}{\mathalpha}{greek}{"22}
\DeclareMathSymbol{\vartheta}{\mathalpha}{greek}{"23}
\DeclareMathSymbol{\varpi}{\mathalpha}{greek}{"24}
\DeclareMathSymbol{\varrho}{\mathalpha}{greek}{"25}
\DeclareMathSymbol{\varsigma}{\mathalpha}{greek}{"26}
\DeclareMathSymbol{\varphi}{\mathalpha}{greek}{"27}

\DeclareSymbolFont{otone}{OT1}{cmr}{m}{n}
\DeclareMathSymbol{\Gamma}{\mathalpha}{otone}{0}
\DeclareMathSymbol{\Delta}{\mathalpha}{otone}{1}
\DeclareMathSymbol{\Theta}{\mathalpha}{otone}{2}
\DeclareMathSymbol{\Lambda}{\mathalpha}{otone}{3}
\DeclareMathSymbol{\Xi}{\mathalpha}{otone}{4}
\DeclareMathSymbol{\Pi}{\mathalpha}{otone}{5}
\DeclareMathSymbol{\Sigma}{\mathalpha}{otone}{6}
\DeclareMathSymbol{\Upsilon}{\mathalpha}{otone}{7}
\DeclareMathSymbol{\Phi}{\mathalpha}{otone}{8}
\DeclareMathSymbol{\Psi}{\mathalpha}{otone}{9}
\DeclareMathSymbol{\Omega}{\mathalpha}{otone}{10}

\DeclareSymbolFont{syms}{OML}{cmm}{m}{it}
\DeclareMathSymbol{\partial}{\mathord}{syms}{"40}

\DeclareMathAlphabet{\mathbold}{OML}{cmm}{b}{it}

\DeclareSymbolFont{largesymbols}{OMX}{cmex}{m}{n}

\newcommand{\commentAA}[1]{}

\title{Hot Streaks on Social Media}

\author{
  Kiran Garimella\thanks{Research done at EPFL.} \\
  MIT\\
  garimell@mit.edu
\And
  Robert West \\
  EPFL\\
  robert.west@epf\/l.ch
}

\begin{document}
\maketitle



\begin{abstract}
Measuring the impact and success of human performance is common in various disciplines, including art, science, and sports.
Quantifying impact also plays a key role on social media, where impact is usually defined as the reach of a user's content as captured by metrics such as the number of views, likes, retweets, or shares.
In this paper, we study entire careers of Twitter users to understand properties of impact.
We show that user impact tends to have certain characteristics:
First, impact is clustered in time, such that the most impactful tweets of a user appear close to each other.
Second, users commonly have ``hot streaks'' of impact, i.e., extended periods of high-impact tweets.
Third, impact tends to gradually build up before, and fall off after, a user's most impactful tweet.
We attempt to explain these characteristics using various properties measured on social media, including the user's network, content, activity, and experience, and find that changes in impact are associated with significant changes in these properties.
Our findings open interesting avenues for future research on virality and influence on social media.
\end{abstract}

\commentAA{
1. "such as number of views, likes, ...". The word "number of" should be there, as you mention that these are metrics.

2. Another high level comment here: perhaps not that relevant to the abstract, but, might be good to add in the introduction to ward off some crazy reviewer critics. Basically, stating somewhere that these metrics are best objective/measurable ways to quantify impact. Ideally, measuring impact can also be subjective based on many other properties of the content and its feedback. For example, papers with high citation counts don't mean they are impactful, as papers with much lower citations could have created a much larger scientific impact, which is true in many real world scenarios. It perhaps might make sense to state this subtle point somewhere in the paper, if you see fit.

3. The phrase "clustered in time" doesn't read that well IMHO. May be "clustered over time" or perhaps even "possess temporal locality" (which might read more naturally to computer scientists) If you choose to make this change, then you will have to also correct all other instances of this statement throughout the paper.

4. Shouldn't you say in the last line: "We 'also' present a simple classifier". Without this it reads a bit weird.
}



\section{Introduction}
\label{sec:introduction}


Evaluating the impact of individual performance is common practice in many realms of life, including art, science, and sports.
Given the importance that society attaches to evaluations of impact---resulting in promotions, awards, and pay raises---,
studying how impact evolves over the course of lives and careers is critical for making sense of human group behavior.
\commentAA{Isn't this statement a bit far-fetched: "It is fair to say our society is built on the evaluation of impact of individuals"?}
Notably, trends in impact over the course of careers have previously been studied
in order to understand the life cycle of creativity~\cite{denisi1981profiles,barrick1991big}, as well as in the field of ``science of science''~\cite{sinatra2016quantifying}.
%





In this paper, we study the impact of users on social media, where impact is usually defined in terms of the reach of a user's content by metrics such as the number of retweets, likes, views, shares, or comments.
More specifically, we focus on \textit{individual impact}---which we define as the reach of an individual user's content 
over the course of their ``career'' (the time since they joined the platform). 
Particularly, we are motivated by recent studies on impact by
\textcite{sinatra2016quantifying}, who show that scientists' most impactful works are randomly distributed over their careers, and
by \textcite{liu2018hot}, who show the existence of periods of increased creativity in scientific, artistic, and cultural careers.
These works lead us to ask the question whether there are similar trends in impact on social media.

Impact is, however, difficult to study, and particularly so in social media, due to the subtle interplay between individual and social aspects of impact.
On the one hand, impact is inherently social: the impact created by a person's content depends on how others in their network perceive the content.
On the other hand, impact is also individual, via factors such activity (number of posts written),
experience (age on the platform), and interests (content of posts).
Due to the challenging nature of the problem, research into long-term trends of individual impact on social media, and the factors that influence these trends, has been limited to date.
Most research has investigated the impact of \textit{isolated pieces of content,} \eg, by predicting retweet counts on Twitter~\cite{suh2010want,martin2016exploring}
or exploring factors that lead to content being shared~\cite{gao2015modeling}.
On the contrary, the literature is much thinner regarding the nature and role of the impact of users over the course of their \textit{entire careers.}

To make progress in this direction, we built four datasets from Twitter, which span over a decade and contain complete careers of users with a wide range of followers, experience, and topical interests.
We choose retweet counts as the measure of impact and reconstruct the careers of users---from their very first post until today---as a time series of retweet counts obtained for their tweets. 
%
We aim to discover characteristic patterns in user impact over time and
to understand
the role played by factors such as the user's network, the content they post, and their levels of activity.
%

\commentAA{While you start the discussion about the most impactful tweets here, there hasn't been any mention till now about the exact metric that has been used in this study to quantify the impact. Is it the retweets, or the views, likes, shares, or a combination of everything? I think it would be good to clarify this before  writing such a statement.}
We start by looking at the position of the most impactful tweets during a user's career and find that, for a large fraction of users, the most impactful tweets are clustered together in time (Section~\ref{sec:clustered_impact}).

Moreover, this trend is not just limited to a few most successful tweets in a user's career, but generalizes to entire periods of high impact (``hot streaks'') during which a user's posts are retweeted significantly more.
We analyze various factors behind this phenomenon, including tweet content, changes in follower counts, retweeting behavior, and activity of the user. 
We find that hot streaks are driven by new retweeters who suddenly start following and retweeting a user, thus leading to a sudden gain in followers during a hot streak. 
Even though these new retweeters retweet highly, their interest is quite limited in time, and they tend to quickly decrease their retweeting activity.
%
Users also become more active during hot streaks, tweeting longer content containing media and a more diverse range of topics (Section~\ref{sec:hot_streaks}).

%
%

Finally, we zoom into the activity around the most retweeted tweets for a user and show that there is a pattern of build-up before, and drop-off after, the most retweeted tweet (Section~\ref{sec:before_after}).

Our study is one of the first to show the presence of such patterns of individual impact on social media.
These findings could help us 
in revisiting existing problems in social network analysis such as understanding virality or influence maximization from an individual perspective.
For instance, the clustering of high-impact tweets and the presence of hot streaks shows when a user is the most impactful in her career. 
This could be useful in finding influential users who are currently on a hot streak to maximize the spread of a marketing campaign. Another example could be to use the presence of the build-up towards a user's most retweeted tweet to explain how tweets become viral for certain users.

\section{Datasets}
\label{sec:datasets}

\subsection{Dataset description}

We use four large datasets with varying characteristics from Twitter. 
%
For each user in our dataset, we obtain all their tweets and construct their \textit{career} as the time series of retweet counts for original tweets posted by the user (retweets by the user are not included in the career).
Each post is associated with a timestamp when it was posted and a measure of impact (retweet count), as measured when the data was crawled (June 2018). 
Below, we briefly describe how each dataset was sampled.
\commentAA{
2. This is the first time when the actual metric to measure the impact is mentioned, i.e., retweet count. Is it a good idea to motivate why retweet count (or a simple citation to support this design choice), or is that an overkill?
3. "Only impact of original content by the user is considered, so retweets are not considered"? Do you mean that the impact of retweets, measured by the retweet counts of retweets is not considered? If it is the case, then isn't retweet counts from retweets already included in the retweet counts of the original tweet? Basically, the motivation of this statement is not clear to me. Or am I missing something?
}

\spara{\verified}
Starting from a list of all verified users on Twitter from June 2018,\footnote{297,000 users, from \url{http://redd.it/8s6nqz}}
we filter out users according to various criteria: they should have between 2,000 and 3,200 tweets, and between 50,000 and 2 million followers.%
\footnote{
The Twitter API gives access to no more than the 3,200 most recent tweets for each user. Since we want to  make sure that we capture the complete career of a user, we restrict our analysis to users with at most 3,200 tweets.
We require at least 2,000 tweets per user because we want to study users with a certain minimum level of activity.
Finally, we remove users with over 2 million followers due to the strict rate limiting constraints on the Twitter API endpoint for getting followers.
}
\commentAA{Grammatical mistakes in footnote number 2. Further even the statement doesn't read that well and is convoluted. Currently it reads like we use 3200 tweets per user because we want to make sure we capture the complete career of a user, but the Twitter API ... .
On the other hand, it should read like: Since twitter API does not provide ..., thus, to consistently reconstruct the career of each user, we restrict ourselves to 3200 tweets.

2000 tweets because we wanted to study users who have a certain activity. Perhaps, "certain minimum level of activity" is better.}

\spara{\political}
This dataset contains over 600,000 politically active users spanning a period of almost 10 years (2009--2018). The dataset has been used in previous work~\cite{garimella2017long}.

\spara{\despoinarandom and \despoina}
We randomly sampled two sets of 1 million users with fewer than 3,200 tweets
from a large sample of 93 million Twitter users collected by~\textcite{antonakaki2018utilizing}. For \despoinarandom, we sampled completely at random, whereas for \despoina we first constrained on certain criteria to ensure high-quality data. In particular, we only selected users who
(i)~have between 2,000 and 3,200 tweets,
(ii)~created their account at least 6 months before April 2016 (time of the crawl by~\textcite{antonakaki2018utilizing}),
and (iii) have at least 100 followers. This gave us 6 million users. 
From this, we randomly sampled 1 million users.

Finally, all datasets were further filtered to retain only those users whose most retweeted tweet received at least 50 retweets.
This was done to ensure that our notion of the most retweeted tweet is meaningful.
For the users who remained after applying these filters, we obtained all tweets, followers, and retweeters of all tweets using the Twitter REST API. 
\commentAA{Kindly rephrase this line, grammatically a bit weird: "For the users left after these filtering". Basically "these filtering" is grammatically wrong, and "for the users left" reads a bit weird.}

Table~\ref{tab:datasets} shows a high-level summary of the datasets, including the number of users, median number of followers, number of tweets, and median of mean retweet count per user (we first compute mean retweet count per user and then take the median over all users). 
The datasets were selected in such a way that they span a wide range of characteristics in terms of content (e.g., politics), user behavior (celebrities vs. normal users), number of followers, and impact.
The \political dataset contains tweets and users related to a specific topic (politics) and might have different behavior due to the 2016 U.S.\ elections and increased political activity;
the \despoina and \despoinarandom datasets capture a random sample of ``normal'' users on Twitter;
and the \verified dataset captures ``celebrity'' behavior on Twitter, as evident from the large median follower count in Table~\ref{tab:datasets}.

In the rest of the paper, 
due to space constraints, whenever the trends are similar on all the datasets, we only show results for the \verified and \despoina datasets. 

Code and datasets used in this paper are publicly available at 
\url{https://github.com/gvrkiran/hot-streaks-social-media}.

\begin{table}[]
\caption{Statistics of our datasets: Number of users, median number of followers, tweets, median of mean retweets and total number of tweets.}
\label{tab:datasets}
{\resizebox{\linewidth}{!}{
\begin{tabular}{l|l|l|l|l|l}
\hline
Dataset             & \#users          & \begin{tabular}[c]{@{}l@{}}med.  \\ \#foll\end{tabular} & \begin{tabular}[c]{@{}l@{}}median  \\ \#tweets\end{tabular} & \begin{tabular}[c]{@{}l@{}}median  \\ \#retweets\end{tabular} & \begin{tabular}[c]{@{}l@{}}total  \\ \#tweets\end{tabular} \\
\hline
\verified       & 2,710     & 105,007 & 2,595.5 &   37.0   & 5.3M        \\
\political       &       10,563     & 360 & 2,353 & 1.52    &        22.3M     \\
\despoina        &       76,777     & 606 & 2,566 & 2.06     &      130M         \\
\begin{tabular}[c]{@{}l@{}}\cpt{Despoina}\\ (\cpt{random})\end{tabular} &       5,465   & 515 &  1,926 &  0.48      &       7.9M       \\
\hline
\end{tabular}}}
\end{table}

\subsection{Historical follower-count estimation}
\label{sec:archive}
%
%
To study the dependence of the follower network on impact, we need historical information on the number of followers a user has at any point in history. However, the official Twitter API does not provide this information. 
We therefore used the Internet Archive Wayback Machine,\footnote{\url{https://archive.org/web/web.php}} which archives web pages 
to obtain historical snapshots of Twitter profiles of users.\footnote{e.g., Barack Obama's Twitter profile has over 7,000 snapshots spanning over 10 years. A snapshot of Barack Obama's Twitter page in 2014 (\url{https://bit.ly/2ssJrkP}) shows that Obama had 46.6M followers at that time.}
We downloaded and parsed the raw HTML from the Internet Archive to obtain follower counts of users at different time stamps.
\commentAA{
1. If I follow the link provided by you in the footnote, I don't see the followers from 2014, however, I see the current set of followers, i.e., 46.6M. Am I missing something here?
https://web.archive.org/web/20140916103521/https:/twitter.com/BarackObama

2. Another general comment. Have you disabled the urls and links in the .tex? Since there are so many links in your paper, perhaps it would be good if these links are clickable so that the reader can directly click on them from the PDF, instead of copying them one by one and pasting in the browser.}

%
%

\begin{figure}[t]
\centering
\begin{minipage}{.46\linewidth}
\centering
\label{}\includegraphics[width=\textwidth, height=\textwidth]{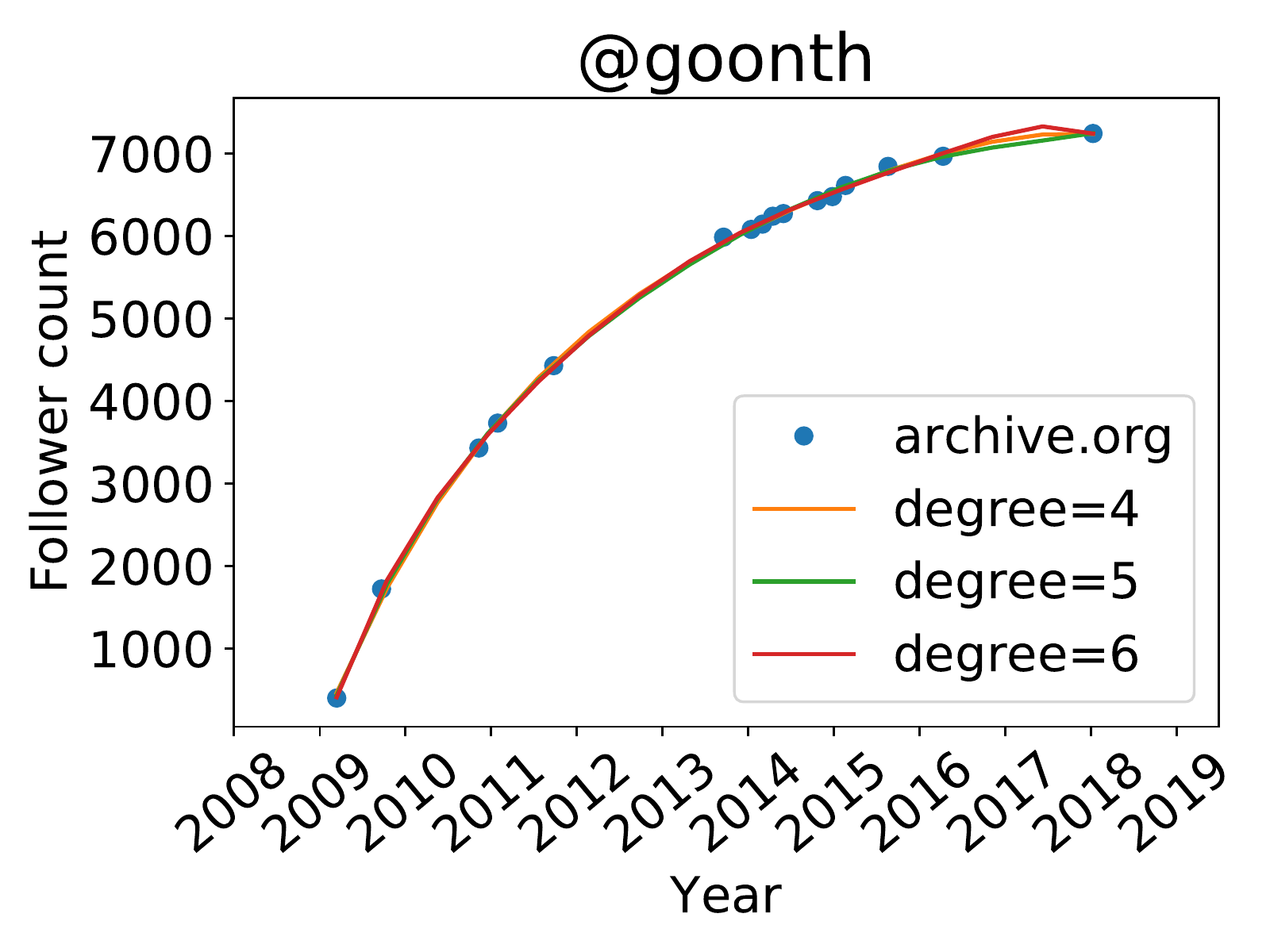}
\end{minipage}%
\begin{minipage}{.46\linewidth}
\centering
\label{}\includegraphics[width=\textwidth, height=\textwidth]{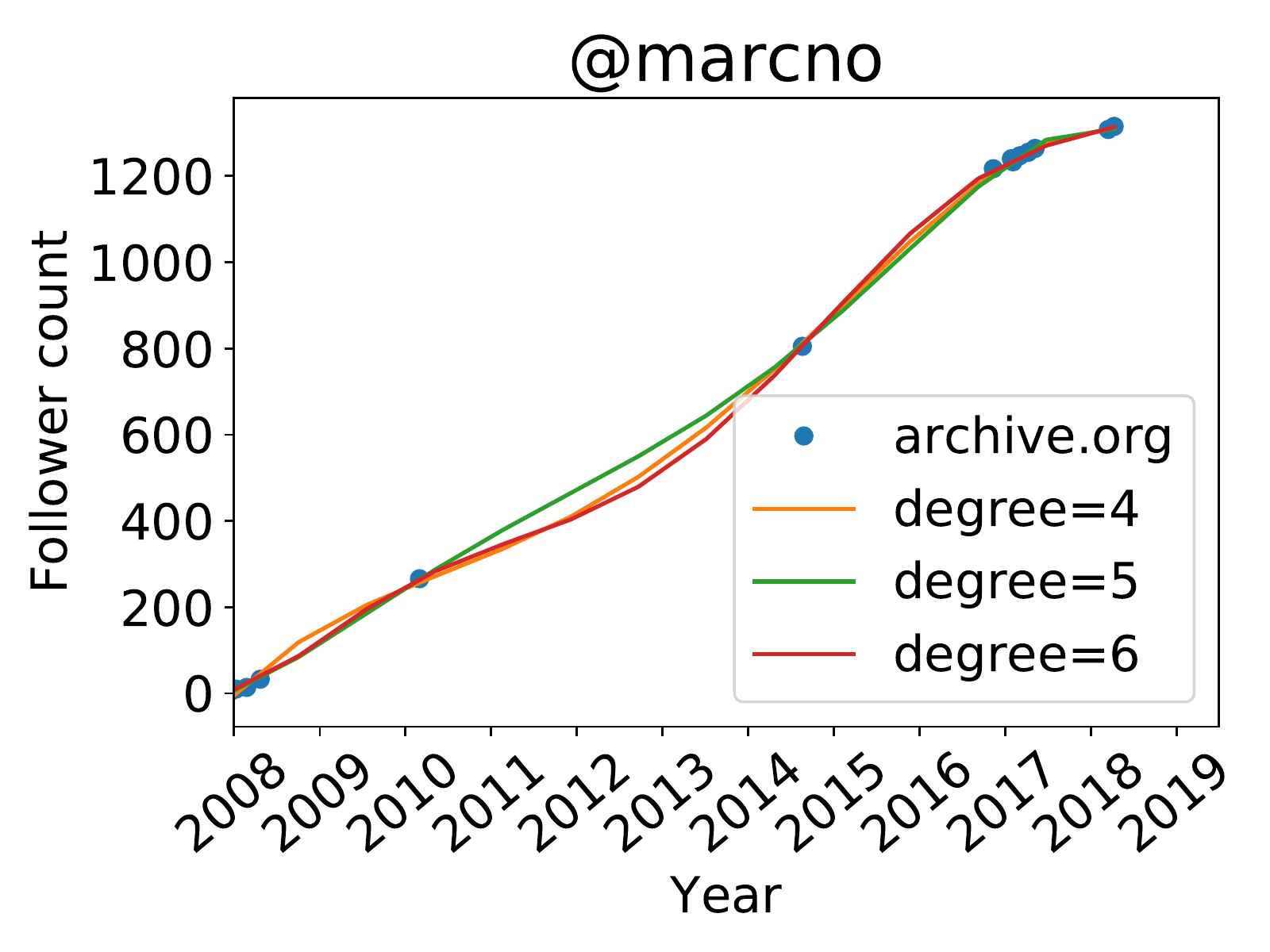}
\end{minipage}%
\commentAA{The font size of the key is too small, and is not visible in Fig. 1. Further, is it also good to add the R2 values for each polynomial fit in Figure 1?}
\caption{Examples of two random userss number of followers over time, obtained from the Internet Archive Wayback Machine (blue dots). The orange, green, and red lines are polynomial fits of degree 4, 5, and 6, respectively.} 
\label{fig:fit_polynomials}
\end{figure}

Not all users have snapshots at all points in time.
We therefore restricted the data to only those users for whom we have at least 10 snapshots, and we
fit polynomial curves to estimate the number of followers for each user at any given point in time.
For each user, we fit polynomials of degrees 4, 5, and 6 and picked the fit with the best $R^2$.
We only consider users for whom the best fit attained $R^2 > 0.9$. Examples of such fits for two random users from our dataset are shown in Figure~\ref{fig:fit_polynomials}. 
We see that the fits capture follower curves of different shapes and provide accurate interpolations for follower counts.
%
\commentAA{"using the best fit polynomial" -> "using the best polynomial fit"}
Now, for each user, using the best polynomial fit, we can estimate the follower count at points in time for which we do not have data from the Internet Archive.
%
%
For instance, for user @marcno in Figure~\ref{fig:fit_polynomials}, we can estimate the number of followers in 2012 as 420.
Using this technique, we obtained the imputed historical number of followers for each user in our dataset. 
All the users included in Table~\ref{tab:datasets} have at least 10 snapshots on the Internet Archive and attain a good fit.

\section{Clustered impact}
\label{sec:clustered_impact}



\commentAA{Do you think there is any explanation required why a week was considered as the temporal resolution for constructing careers?}
We start with some notation.
All our analysis will be at a user level, and we aggregate over all users.
Assume a user $u$ has $N$ tweets, $t_1$, \dots, $t_N$.
We call the life time of a social media user, from their first tweet to their last tweet (at data collection time, June 2018), the \textit{career} of that user.
%
The position of a tweet $t_i$ in a user's career is measured in terms of
(i)~time in weeks since the first tweet by the user, $w(t_i)$
and (ii)~tweet index $i$, given by $i = P(t_i)$.
Though time and tweet index might be correlated, there are users with long periods of inactivity, for whom using absolute weeks to measure position would be misleading. Hence, we perform all our experiments using both measures. Unless explicitly specified, we use tweet index to denote the position of a tweet.
The $k$ most retweeted tweets in a user's career are denoted by $T_{1}, T_2, T_3,$ \dots, $T_k$.

\begin{figure}[t]
\centering
\begin{minipage}{.99\linewidth}
\centering
\subfloat[\verified]{\label{}\includegraphics[width=\textwidth, height=0.25\textwidth]{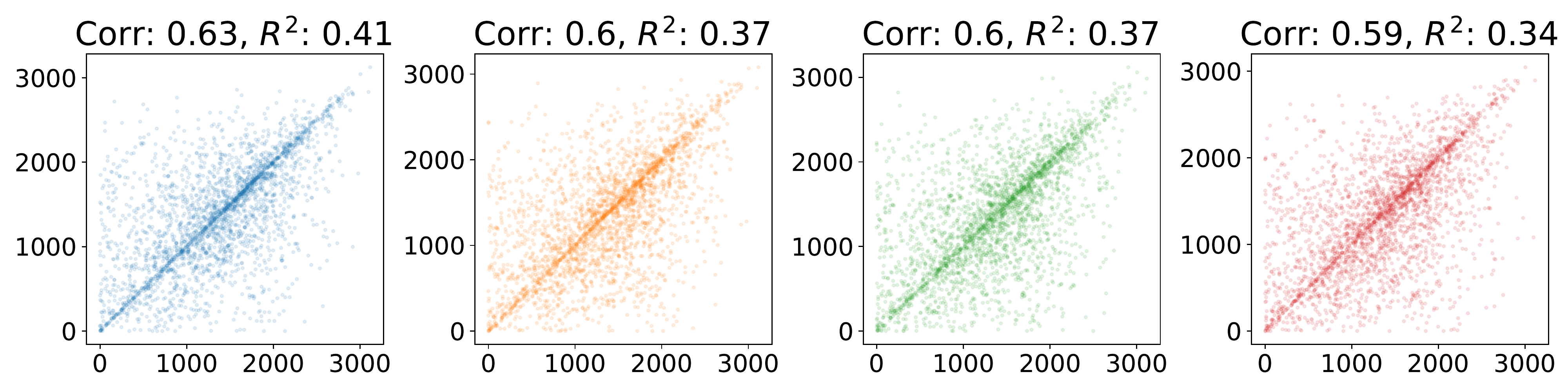}}
\end{minipage}%
\par\medskip
\begin{minipage}{.99\linewidth}
\centering
\subfloat[\political]{\label{}\includegraphics[width=\textwidth, height=0.25\textwidth]{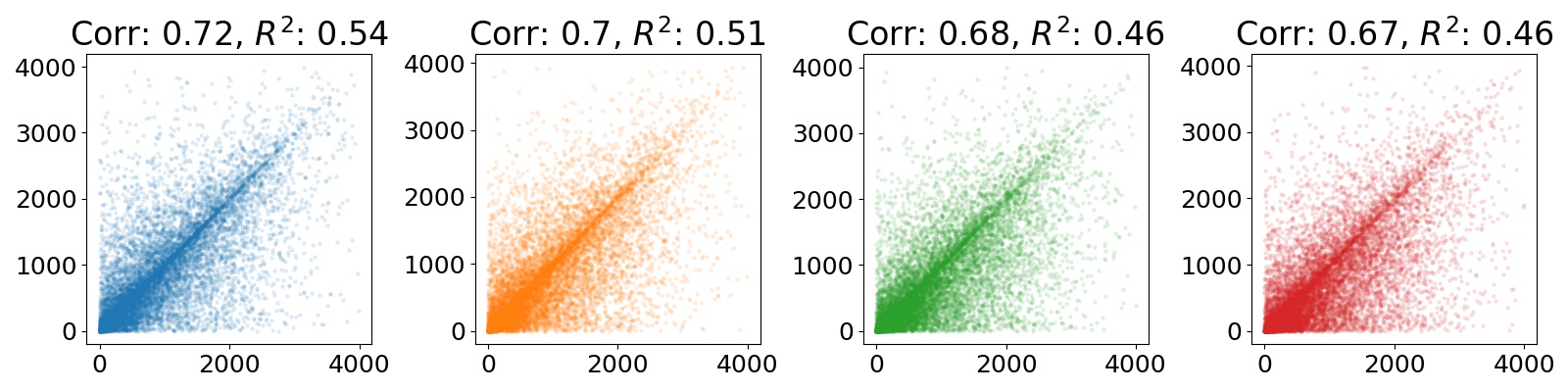}}
\end{minipage}%
\par\medskip
\begin{minipage}{.99\linewidth}
\centering
\subfloat[\despoina]{\label{}\includegraphics[width=\textwidth, height=0.25\textwidth]{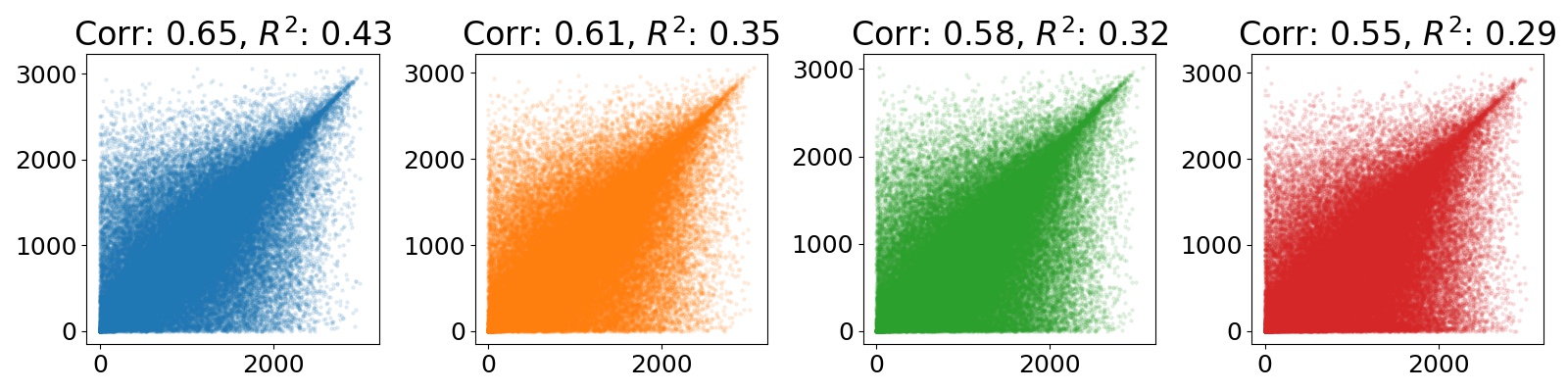}}
\end{minipage}%
\par\medskip
\begin{minipage}{.99\linewidth}
\centering
\subfloat[\despoinarandom]{\label{}\includegraphics[width=\textwidth, height=0.25\textwidth]{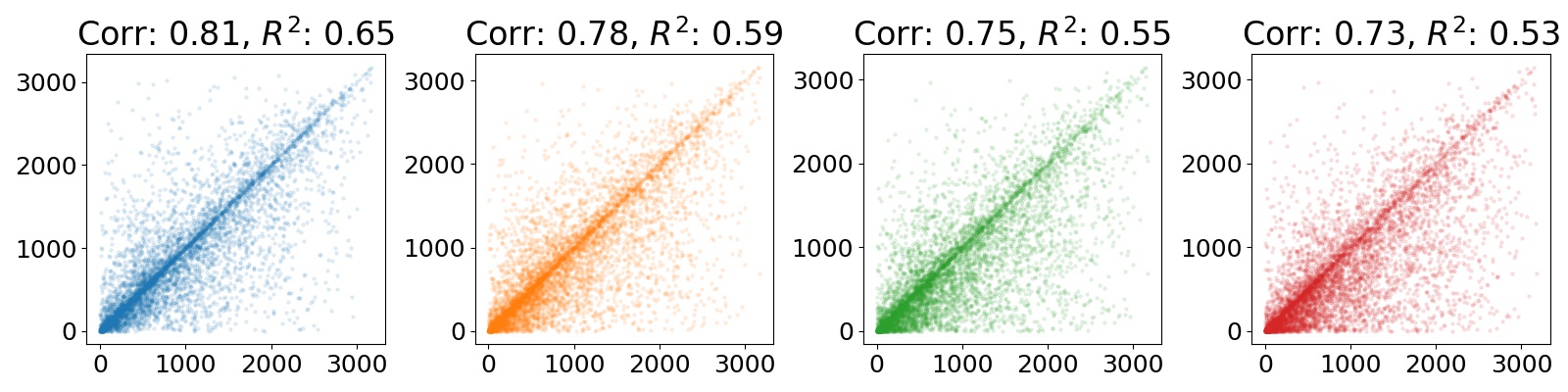}}
\end{minipage}%
\caption{Scatter plots showing the correlation of the tweet position of the five most retweeted tweets in a user's career. Each row shows the correlation between the position of the most retweeted tweet ($P(T_1)$) and that of the $i$-th most retweeted tweet ($P(T_{i})$), for $i \in [2,5]$ (left to right). Each dot represents a user. We see that most users lie close to the diagonal, indicating that the most retweeted tweets occur closely in time.
``Corr.''\ denotes Pearson's correlation coefficient.
Note that in \verified and \despoina, all users have between 2,000 and 3,200 tweets.
}
\label{fig:hotstreaks_scatter}
\end{figure}

We begin by investigating the timing of the five most retweeted tweets in a user's career, $T_i$ for $i \in [1,5]$.
Scatter plots of $P(T_1)$ \vs $P(T_i)$ for $i \in [2,5]$ are shown in Figure~\ref{fig:hotstreaks_scatter}.
We observe that the positions of $T_1$ and $T_{i}$ are highly correlated.
We also computed all pairwise correlations between $P(T_i)$ and $P(T_{i+1})$, and between $w(T_i)$ and $w(T_{i+1})$ (for $i \in [1,4]$), 
and obtained similar results. 
To check the robustness of these results, we shuffled the order of the tweets in each career, and the correlation drops from around 0.6 to below 0.10 for all datasets, indicating that this phenomenon holds strongly for real careers.

\begin{figure}[t]
\centering
\includegraphics[width=0.32\textwidth, height=0.32\textwidth]{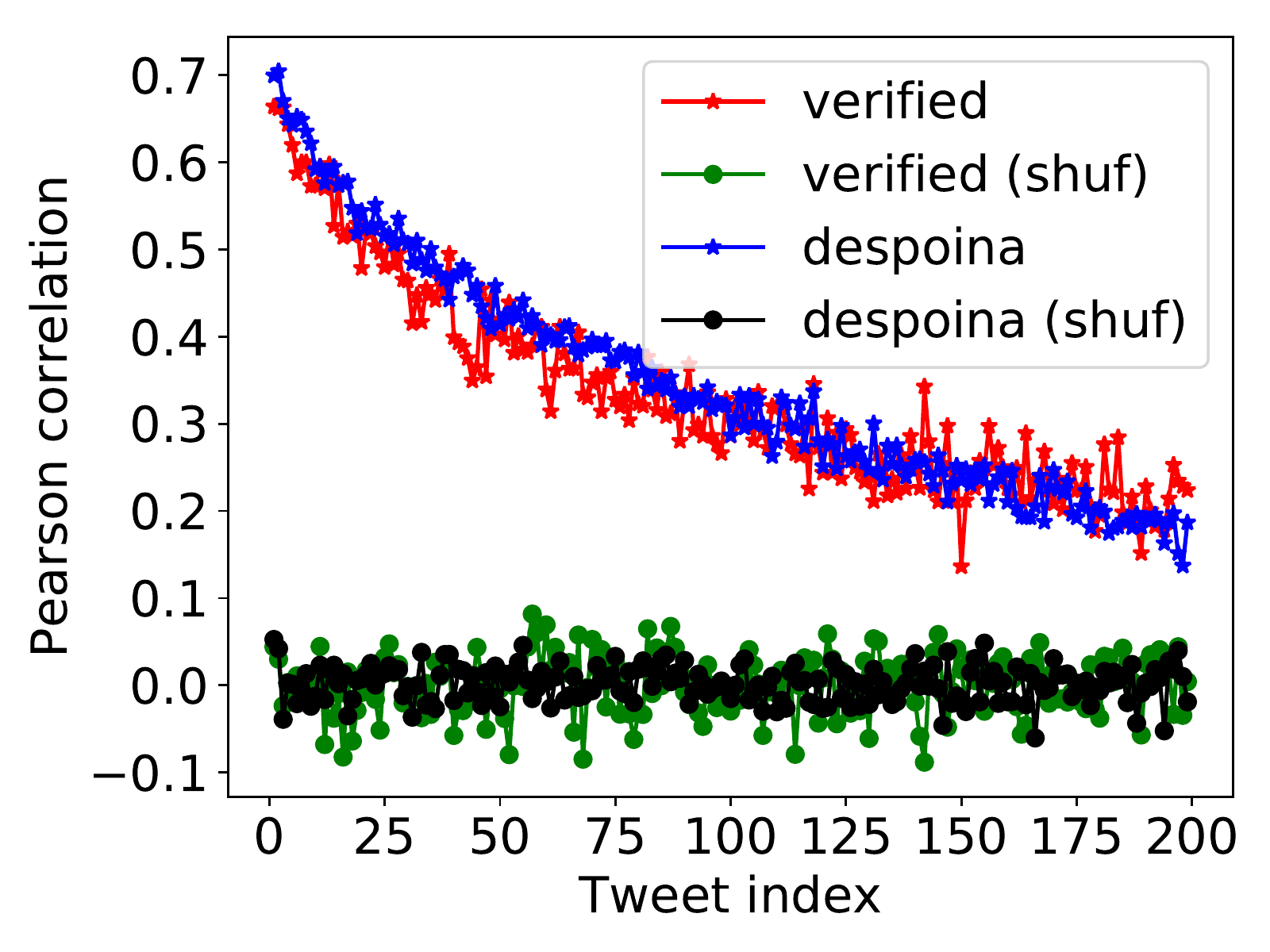}
\caption{
Pearson correlation for $T^*_1$ and $T^*_i$ for $i \in [2,200]$.}
\label{fig:correlation}
\end{figure}

A user's retweet counts grow over their career, 
 in part due to the growing number of followers a user attracts over her career~\cite{suh2010want,jenders2013analyzing}.
 Hence, the likelihood of a user having their most retweeted tweets later in their career is higher \textit{a priori}.
To mitigate such effects, we normalize retweet counts at a point in time by the number of estimated followers at that time (\cf Section~\ref{sec:archive}), computing retweets per follower for every tweet in a user's career. 
We denote the $k$ tweets with the most retweets per follower by $T^*_{1}, T^*_2, T^*_3,$ \dots, $T^*_k$.
In order to check how well the clustering in Figure~\ref{fig:hotstreaks_scatter} goes beyond the the top 5 most retweeted tweets, we computed the correlation between the position of the tweet with the most retweets per follower, $P(T^*_{1})$, with $P(T^*_{2})$, \dots, $P(T^*_{200})$.
The results for the \verified and \despoina datasets are shown in Figure~\ref{fig:correlation}, where each point represents the Pearson correlation between $T^*_1$ and $T^*_i$ for $i \in [2,200]$.
We see that the correlation remains above 0.5 for up to 25 tweets.
In Figure~\ref{fig:correlation}, we also show the correlation for shuffled careers (green and black lines) and we can see that real careers have a much higher correlation than shuffled careers.

\begin{figure}[t]
\begin{minipage}{.49\linewidth}
\includegraphics[width=\textwidth, height=\textwidth, clip=true, trim=0 0 0 0]{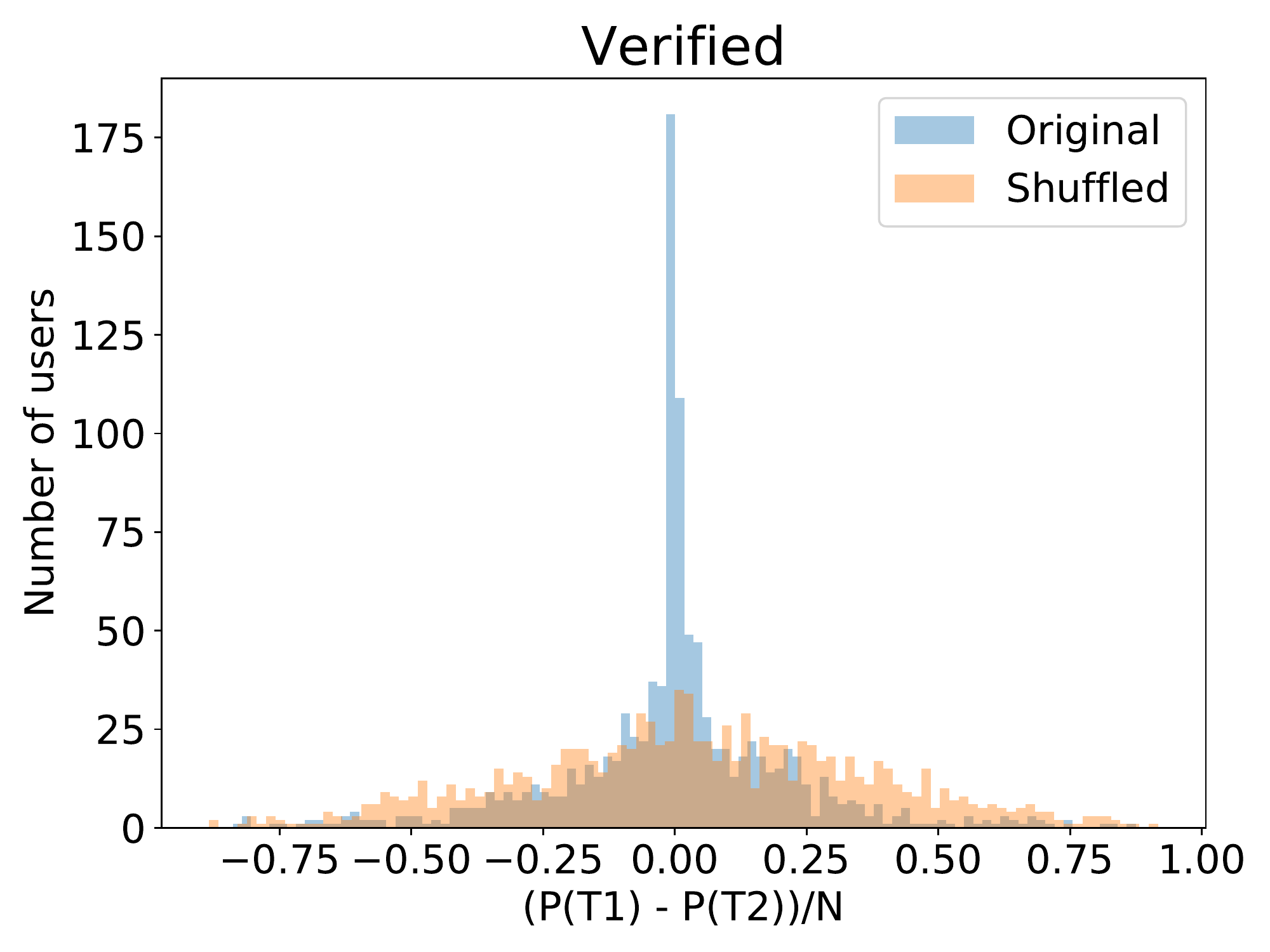}
\end{minipage}%
\begin{minipage}{.49\linewidth}
\includegraphics[width=\textwidth, height=\textwidth, clip=true, trim=0 0 0 0]{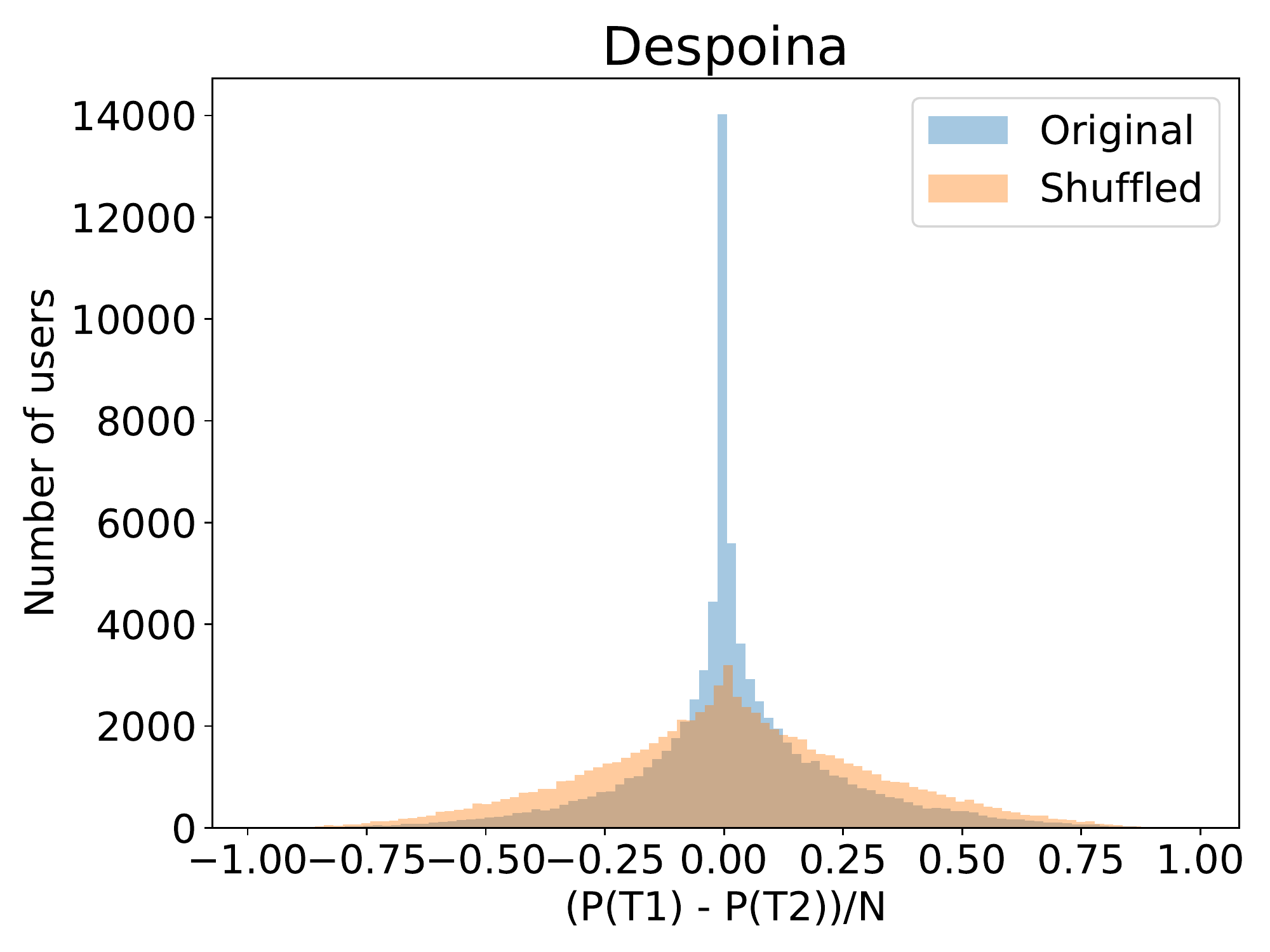}
\end{minipage}%
\caption{Normalized difference between the position of the first and second most retweeted tweets.}
\label{fig:position_difference}
\end{figure}

Next, we look at the difference in the position of the top two most retweeted tweets and compute the normalized difference $(P(T_1) - P(T_2))/N$ for each user.
Figure~\ref{fig:position_difference} shows the distribution of the difference. 
We see that the distribution peaks and is centered at zero, providing further evidence that $T_1$ and $T_2$ are close to each other. The fact that the distribution is symmetric around zero indicates that $P(T_1)$ can be before or after $P(T_2)$. 
We compared this with a distribution of the difference for shuffled careers (orange bars in Figure~\ref{fig:position_difference}) and observe that the strong peak close to zero disappears, as expected. 
Similar patterns hold for the other datasets and when using $(P(T_i)-P(T_{i+1}))/N$, for $i \in [2,4]$, instead of the above-defined difference. 



\section{Existence of hot streaks}
\label{sec:hot_streaks}

In this section, we try to generalize the finding from the previous section to see if such highly retweeted tweets
occur within extended periods of higher-than-usual-impact tweets.
We call such periods ``hot streaks'' following its usage in recent work~\cite{liu2018hot}.

%
\subsection{Identifying hot streaks}
\label{sec:identifying}
We propose a simple method to identify hot streaks. We represent the retweets a user gets throughout her career as a time series and identify periods of consistently high activity in this time series. 
%
We identify such periods in a user's career by solving an optimization problem.
The objective is to minimize the least-squares fit of a piecewise\hyp constant function to the retweet counts time series and observe if there are extended periods with a high constant. 
To avoid overfitting, we add a regularization term on the number of constant functions used.

Specifically, the problem can be defined as such: 
given a user $u$ whose career consists of $N$ tweets with retweet counts $r_1$, \dots, $r_N$, 
find $N$ constants $c_1$, \dots, $c_N$ that
\begin{equation}
\label{eq:optimization}
    \text{minimize} \;\;\;
    \sum_{i=1}^{N} (c_i - r_i)^2  + \alpha \sum_{i=1}^{N-1}\delta(c_i,c_{i+1}),
\end{equation}
where $\delta(a,b)=0$ if $a=b$, and $\delta(a,b)=1$ otherwise.
%

\begin{figure}[t]
\begin{minipage}{.49\linewidth}
\includegraphics[width=\textwidth, height=\textwidth]{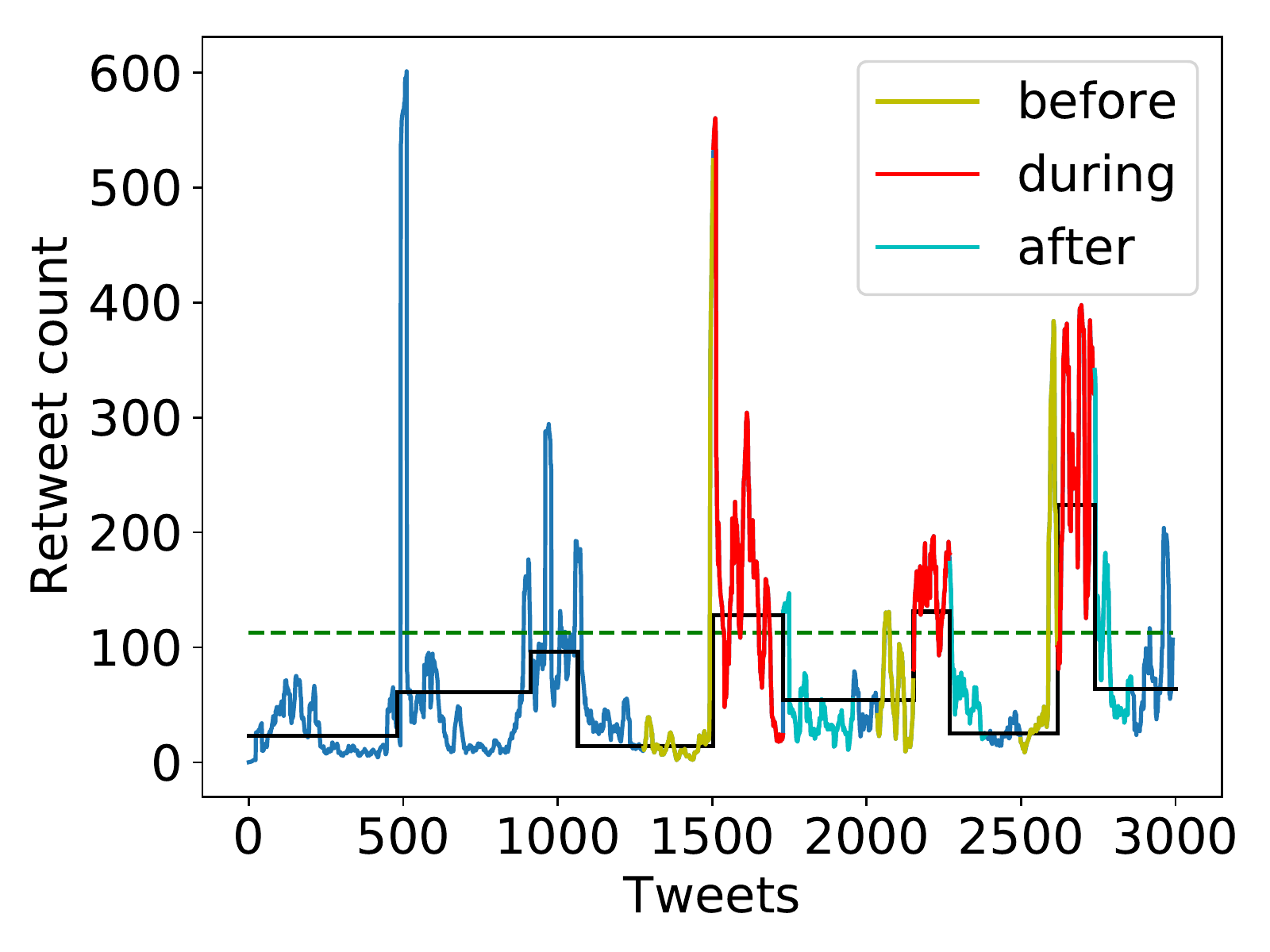}
\end{minipage}%
%
\begin{minipage}{.49\linewidth}
\includegraphics[width=\textwidth, height=\textwidth]{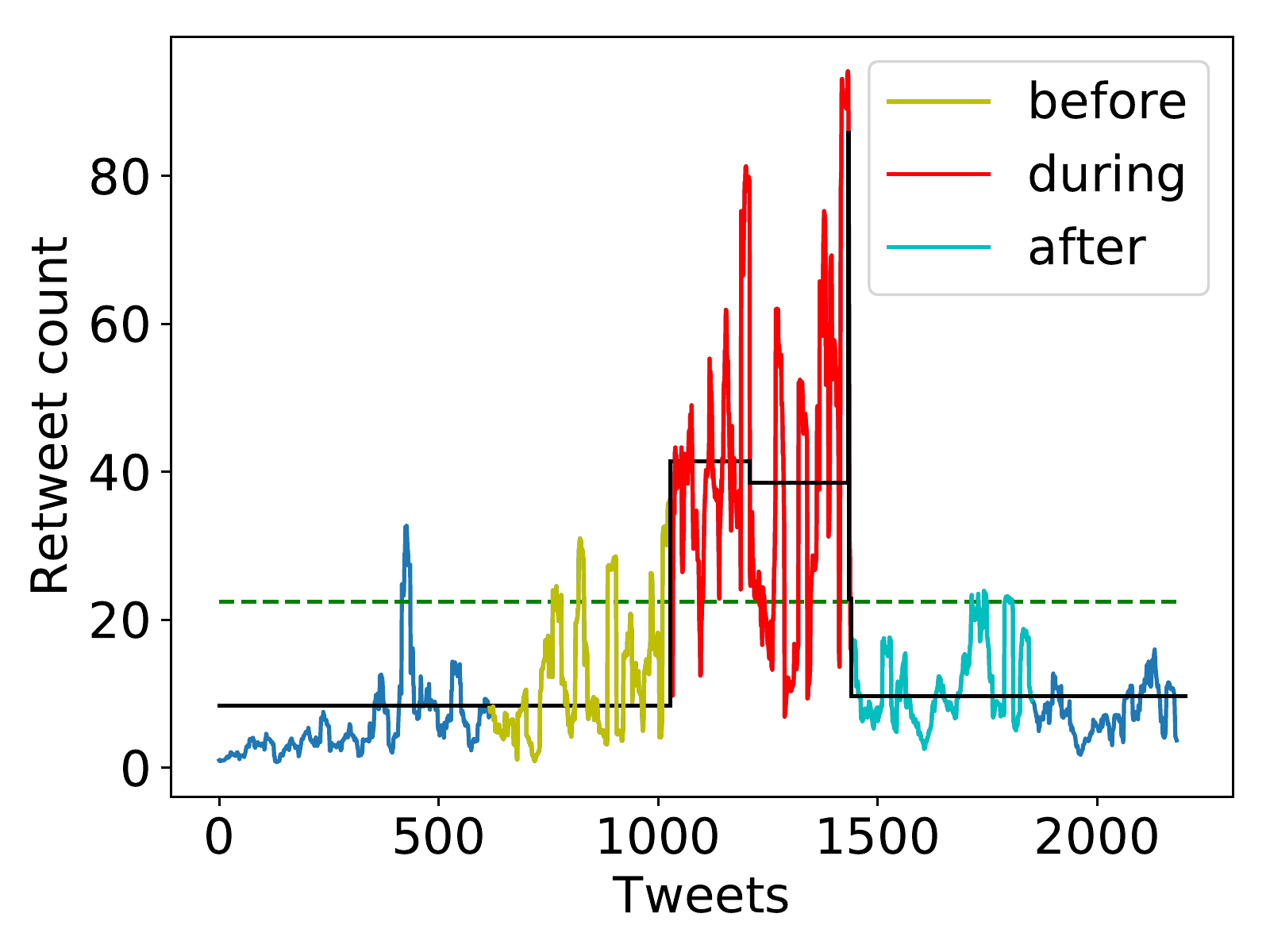}
\end{minipage}%
\par\medskip
\caption{Example careers of two users. The blue line shows the retweet counts over the career of a user. The black line shows the piecewise\hyp constant fit obtained using Equation~\ref{eq:optimization}. The dashed green horizontal line shows the 90th percentile. The periods of increased activity indicating hot streaks can be seen in red. Periods before and after the hot streak are labeled in yellow and cyan, respectively.}
\label{fig:hotstreaks_examples}
\end{figure}

The objective function in Equation~\ref{eq:optimization} tries to balance between too many constants (first term) and too few (second term).
The penalty parameter $\alpha$ controls the number of constants.
There are exponentially many functions (sets of constants) that would need to be evaluated when solving Equation~\ref{eq:optimization} by brute force. We can, however, also solve it efficiently using a dynamic programming solution with time complexity $O(n^2)$~\cite{wang2011ckmeans}.
Examples of two users' careers with fits obtained using Equation~\ref{eq:optimization} are shown in Figure~\ref{fig:hotstreaks_examples}.
The blue line shows the raw retweet\hyp count time series;
the black lines show the piecewise\hyp constant fit.

%

Given these piecewise\hyp constant fits, we define a period as a hot streak if its constants lie consistently above a certain threshold, say, the $k$-th percentile of retweets in a users career.
We also experimented with other methods,%
\footnote{
We tried
(i)~identifying contiguous portions of a career above a certain threshold,
(ii)~using time-series change-point detection~\cite{liu2013change,ross2015parametric},
and (3)~Fourier analysis~\cite{moskvina2003algorithm}.
}
but found that these algorithms were not robust enough to handle the spiky nature of retweets.
For instance, we see in Figure~\ref{fig:hotstreaks_examples} (right) that there are many tweets during the hot streak (marked in red) below the threshold (dashed green line). Using these simple methods fails to identify such cases. Note that the exact method used to identify hot streaks is orthogonal to the analysis in the rest of the paper. Hence, any other method can also be used.


\spara{Parameters} We have two parameters in the optimization problem for identifying hot streaks:
(i)~$\alpha$ in Equation~\ref{eq:optimization}: we tested for multiple values of $\alpha$, from 1 to 10 and fixed the value of 1 because it gave the best fits (observed via manual inspection);
(ii)~the percentile threshold $k$ to identify a hot streak:
we tested a range of values from 70 to 95 in increments of 5 and fixed the value of $k=90$ (i.e., 90th percentile); higher\slash lower values lead to fewer\slash more hot streaks.
%

\subsection{Characteristics of hot streaks}
\label{sec:characteristics}
The real value of identifying hot streaks is in revealing several fundamental patterns that govern individual careers.
In this section, we look at various characteristics of hot streaks.
Similar to the previous section, for every analysis we perform, we compared the results with shuffled careers, thus ensuring the robustness of our results.

\begin{figure}
\begin{minipage}{.49\linewidth}
\includegraphics[width=\textwidth, height=\textwidth]{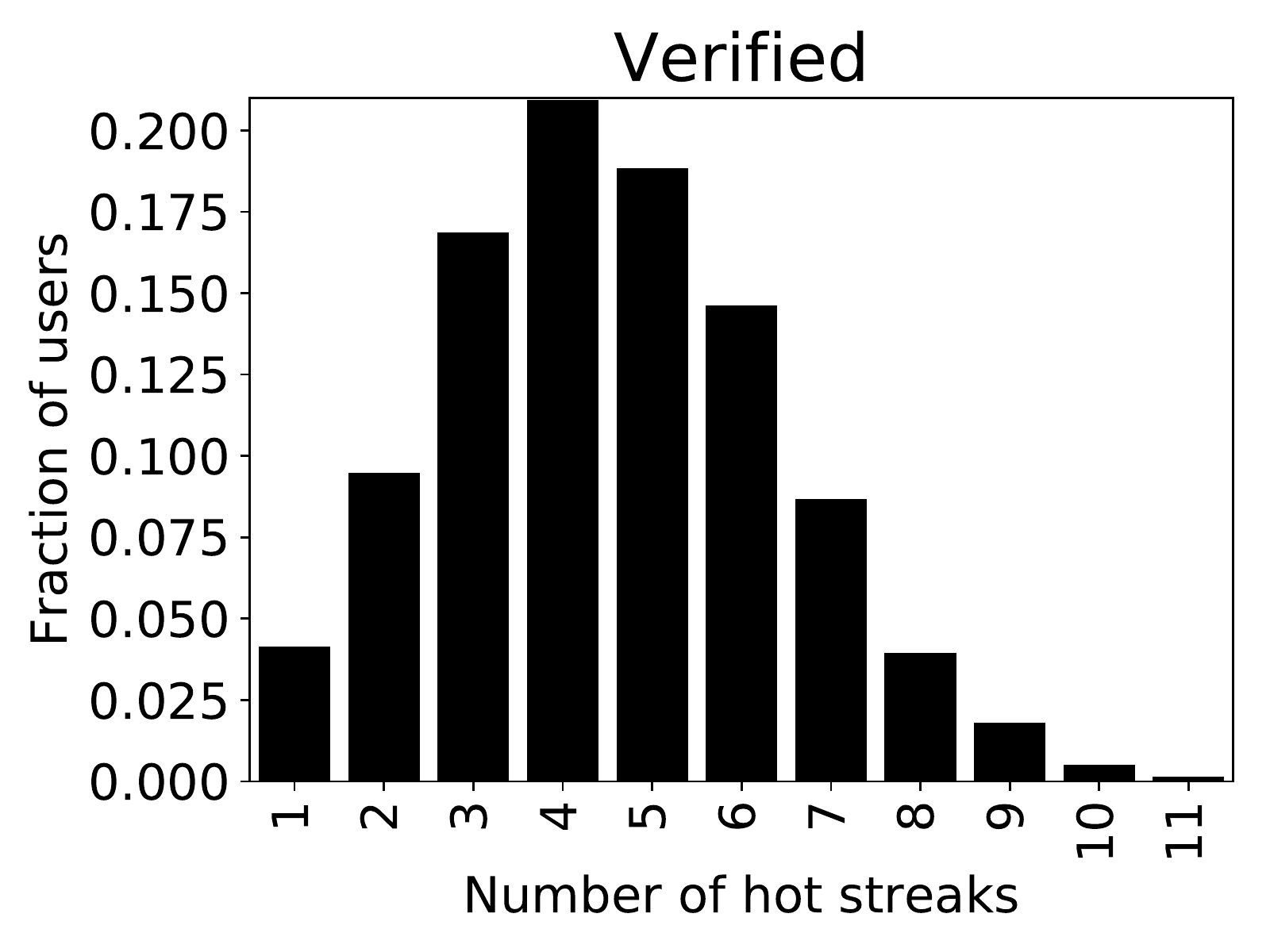}
\end{minipage}%
\begin{minipage}{.49\linewidth}
\includegraphics[width=\textwidth, height=\textwidth]{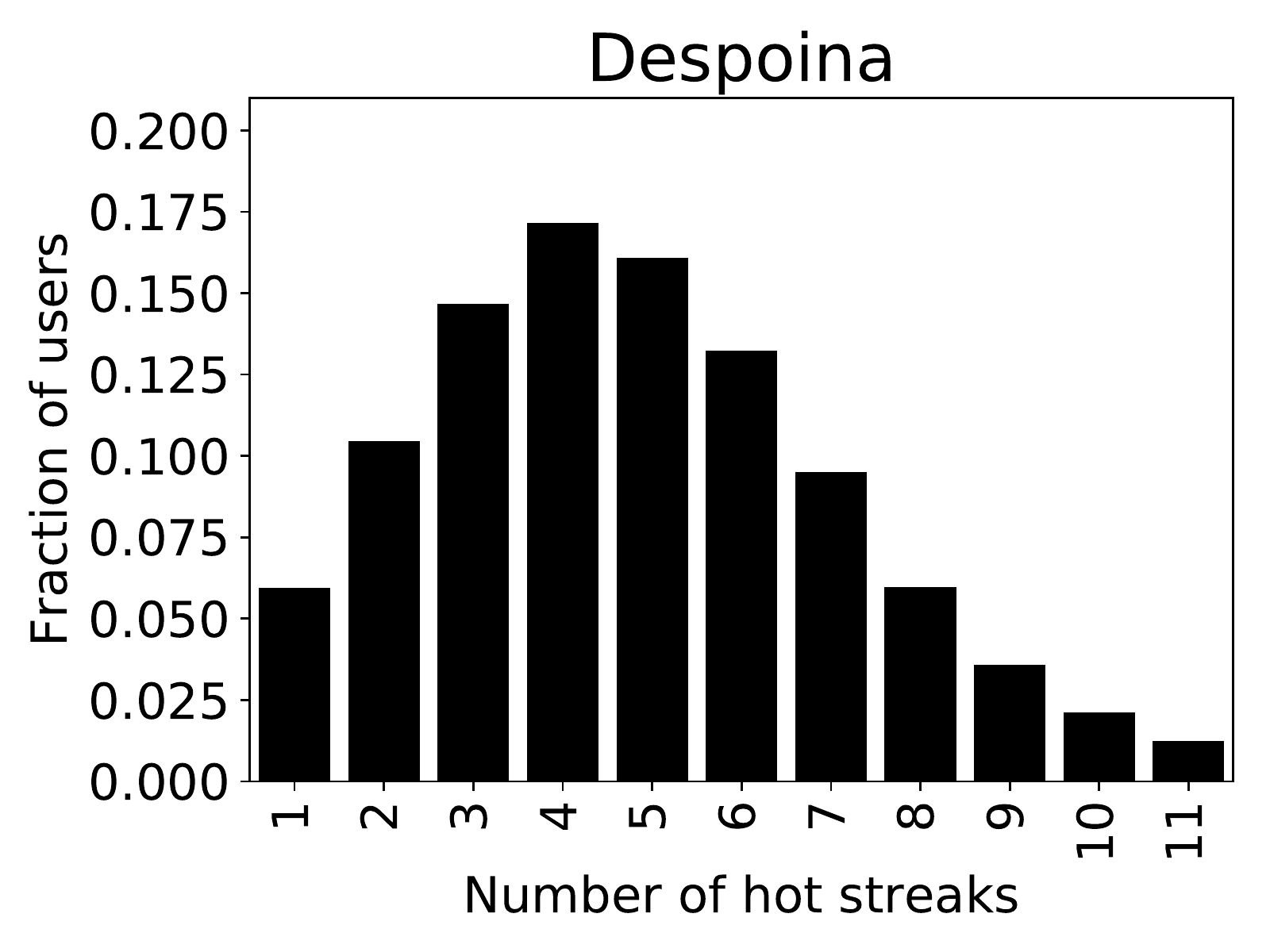}
\end{minipage}%
\caption{Distribution of number of hot streaks.}
\label{fig:hotstreaks_characteristics}
\end{figure}

\spara{Number of hot streaks} We first look at the fraction of users for whom a hot streak greater in length than 10 tweets exists. 
We find that 86\% of users in \verified, 59\% in \political, 63\% in \despoina, and 65\% in \despoinarandom have at least one such hot streak.
We compared these with the fraction of users having at least one hot streak for randomly shuffled careers, and the fractions drop to less than 30\%.
Given that each of these datasets was sampled in a different way and that a nontrivial fraction of users have hot streaks, we can conclude that hot streaks are a prevalent phenomenon.
Figure~\ref{fig:hotstreaks_characteristics}, shows the distribution of the number of hot streaks per use. We see that a majority of users have fewer than 4 hot streaks, with some users  having more than 10.

\begin{figure}
\begin{minipage}{.49\linewidth}
\includegraphics[width=\textwidth, height=\textwidth]{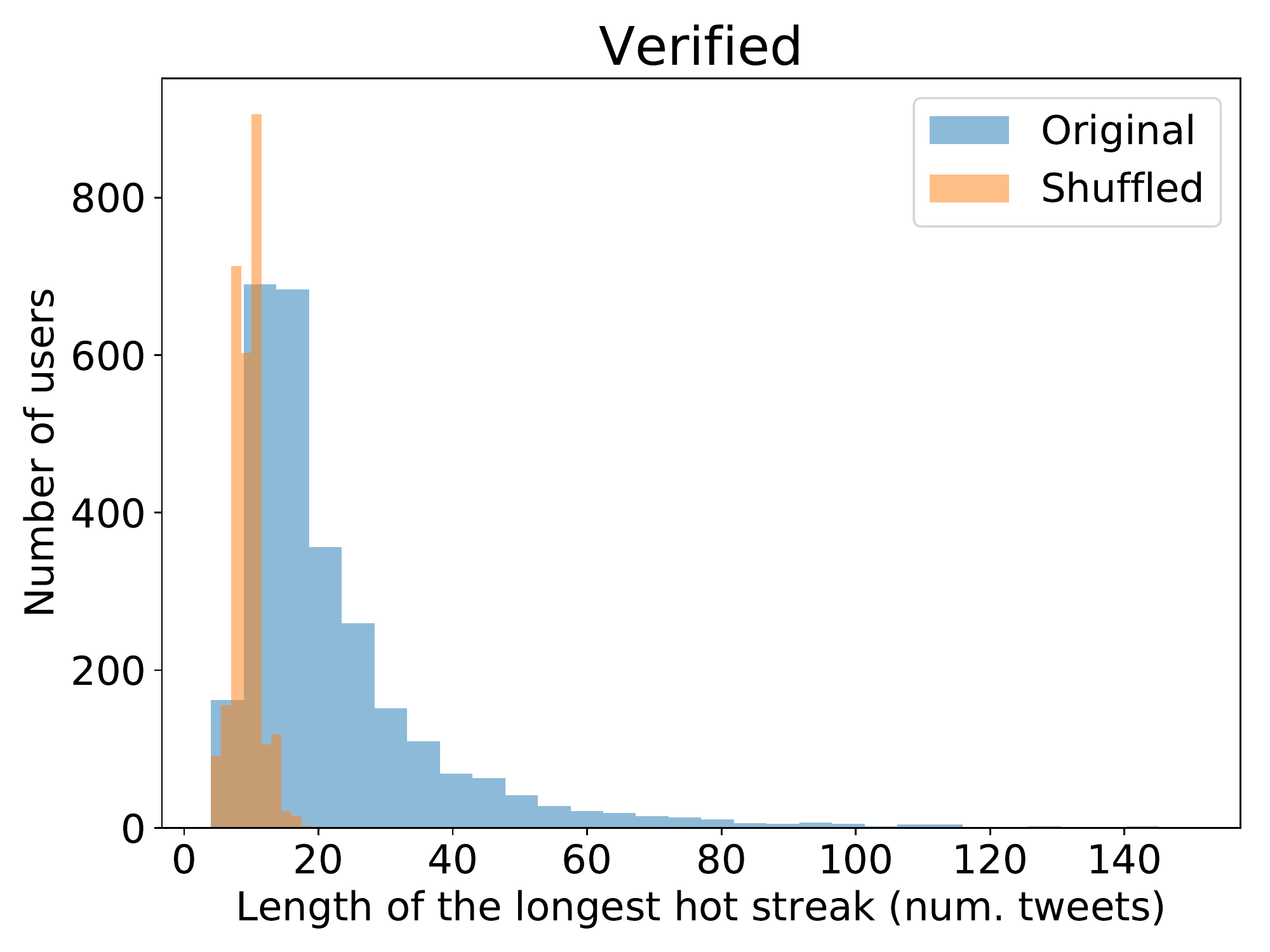}
\end{minipage}%
\begin{minipage}{.49\linewidth}
\includegraphics[width=\textwidth, height=\textwidth]{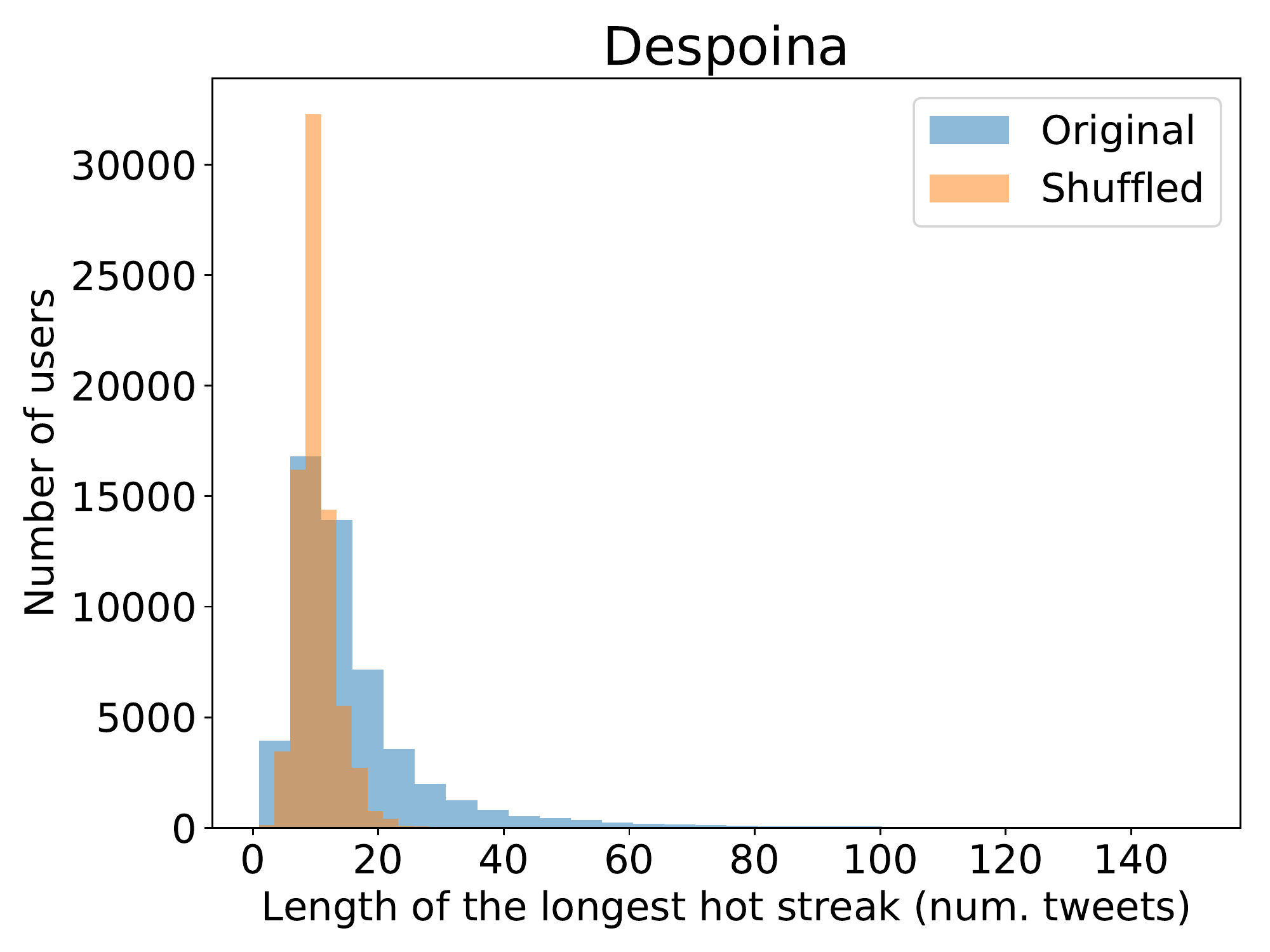}
\end{minipage}%
\caption{Distribution of length of hot streaks compared to shuffled careers.}
\label{fig:hotstreaks_length}
\end{figure}

\spara{Length of hot streaks} Next, we compute the distribution of the length of the longest hot streak for all users. We can see from Figure~\ref{fig:hotstreaks_length} that for a majority of users, hot streaks are mostly short in length---up to 20 tweets---, but some users have much longer hot streaks of over 100 tweets.
%
We compare the lengths of hot streaks to shuffled careers and 
see in Figure~\ref{fig:hotstreaks_length} that real careers are characterized by a longer tail indicating that real users have longer streaks of high retweet activity.
We tested the robustness of these results by controlling for individual career length and arrived at the same conclusions.

\begin{figure}
    \centering
    \includegraphics[width=0.5\textwidth, height=0.15\textwidth]{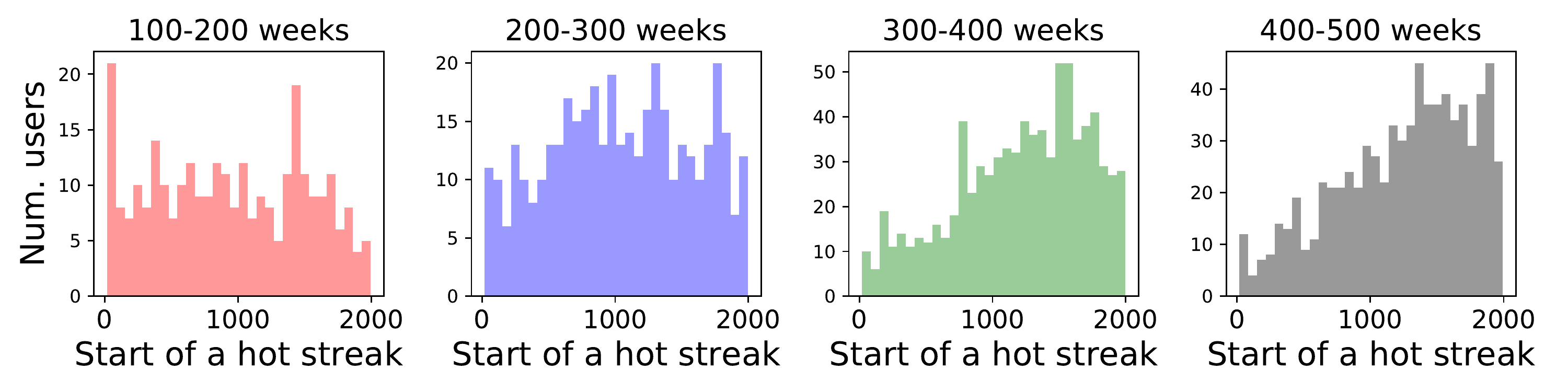}
    \caption{Distribution of the position of hot streaks in user careers (\verified dataset), for various user ages on Twitter (one plot per age). Only the first hot streak of each user is considered.}
    \label{fig:hotstreaks_position}
\end{figure}

\spara{Position of hot streaks} What is the position of hot streaks in a user's career? Are they randomly distributed, as found by~\textcite{liu2018hot} in the context of art, science, and movies?
Defining the position of a hot streak as the position of its first tweet (within the respective user's career), we find that the position of hot streaks in a user's career differs depending on how long the user has been on Twitter (which we refer to as the ``age on Twitter''). 
We divided users into 4 categories based on their age on Twitter: 100--200, 200--300, 300--400, and 400--500 weeks.\footnote{(i) Since Twitter introduced the retweet button only in 2009, we only consider users after that. Users with a career over 500 weeks would have their first tweet prior to 2009 and hence were not considered. (ii) The results remain similar for different groupings, \eg, when bucketing into groups spanning 50, rather than 100, weeks.}
Figure~\ref{fig:hotstreaks_position} shows the distribution of the position of hot streaks in careers.
We see that it is positioned at different places depending on the user's age on Twitter. 
%
For older users, it is more likely to be at the end of their career. As users spend more time on the platform, they learn more, gain more followers, and hence their chance of having a hot streak increases.
For younger users, the position is more evenly distributed, indicating other factors at play (content, activity, etc.).

\spara{User properties} Who are the users who have hot streaks? How are they different from the others who do not have hot streaks? To check if users who have hot streaks differ in any way from those who do not, we compared the two user groups in terms of profile features such as the number of followers, friends, tweets, and age on Twitter. We find that users with hot streaks have a significantly higher number of followers, but that there is no difference in the other feature values. 
%

\begin{figure}[ht]
\centering
\begin{minipage}{.49\linewidth}
\centering
\includegraphics[width=0.9\textwidth, height=0.9\textwidth]{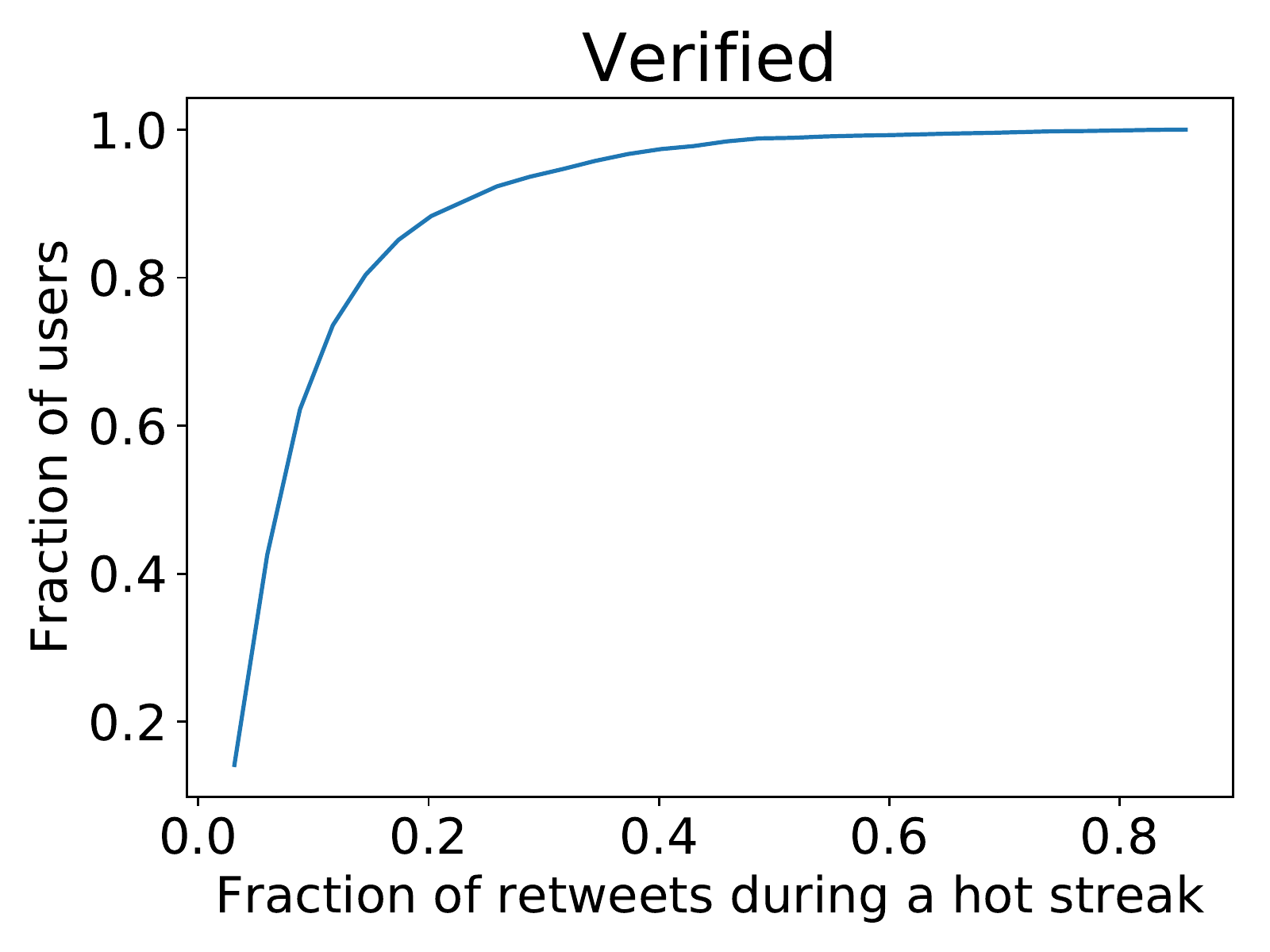}
\end{minipage}%
\begin{minipage}{.49\linewidth}
\centering
\includegraphics[width=0.9\textwidth, height=0.9\textwidth]{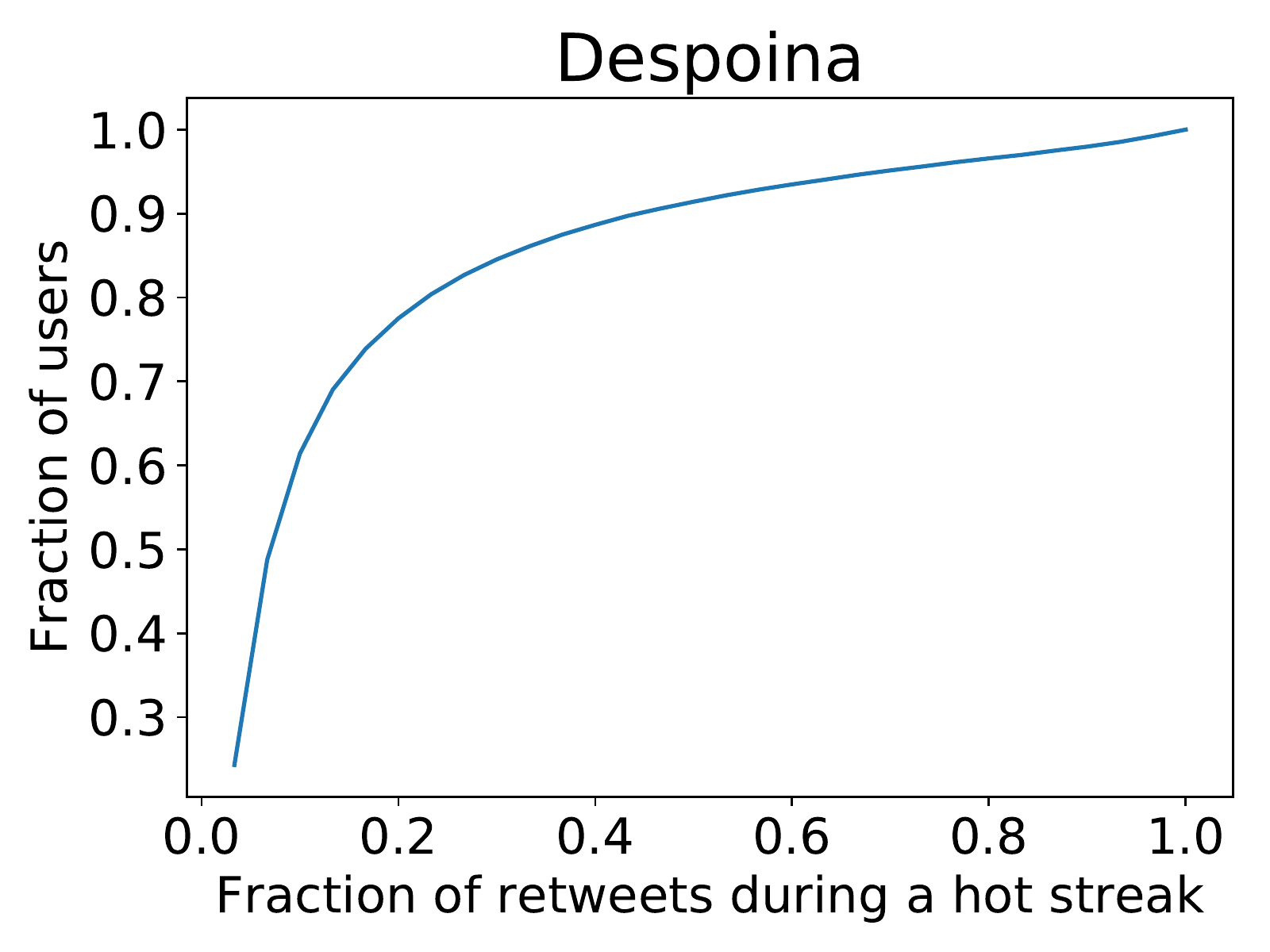}
\end{minipage}%
\caption{CDF of the fraction of retweets obtained during a hot streak.}
\label{fig:rt_count_frac}
\end{figure}

\spara{Influence} How influential are hot streaks in defining success in a user's career? To answer this question, we compute the fraction of total retweets obtained during all hot streaks in a user's career. Figure~\ref{fig:rt_count_frac} shows the distribution of the retweet count fraction for two datasets. We see that for a majority of users (above 80\%), only 10--20\% of retweets are obtained during a hot streak, indicating that hot streaks are not the main driver of retweet activity  in a user's career.

Finally, we investigate whether hot streaks are connected to a recently introduced feature on Twitter--- threads\footnote{\url{https://www.wired.com/story/twitter-multi-tweet-threads/}}---by looking at hot streaks for a random sample of 50 users manually.
In only one of the 50 cases did the hot streak occur in a thread, indicating that hot streaks are typically not threaded tweets.

\subsection{What happens during a hot streak?}

In the analysis below, for each user, we only consider the longest hot streak and obtain tweets from periods of equal length (in terms of number of tweets) before and after the hot streak. That is, if a user has a hot streak of length $k$ tweets, we consider a period of $k$ tweets before and after the hot streak (examples are shown in Figure~\ref{fig:hotstreaks_examples}, colored in yellow and cyan). All the results reported are statistically significant ($p< 0.01$) after applying Bonferroni correction for multiple tests, wherever applicable.
All the results presented below are consistent across datasets, but we focus on the \verified dataset due to space constraints.

\begin{figure*}
\centering
\begin{minipage}{.24\linewidth}
\centering
\subfloat[]{\label{}
\includegraphics[width=\textwidth, height=\textwidth]{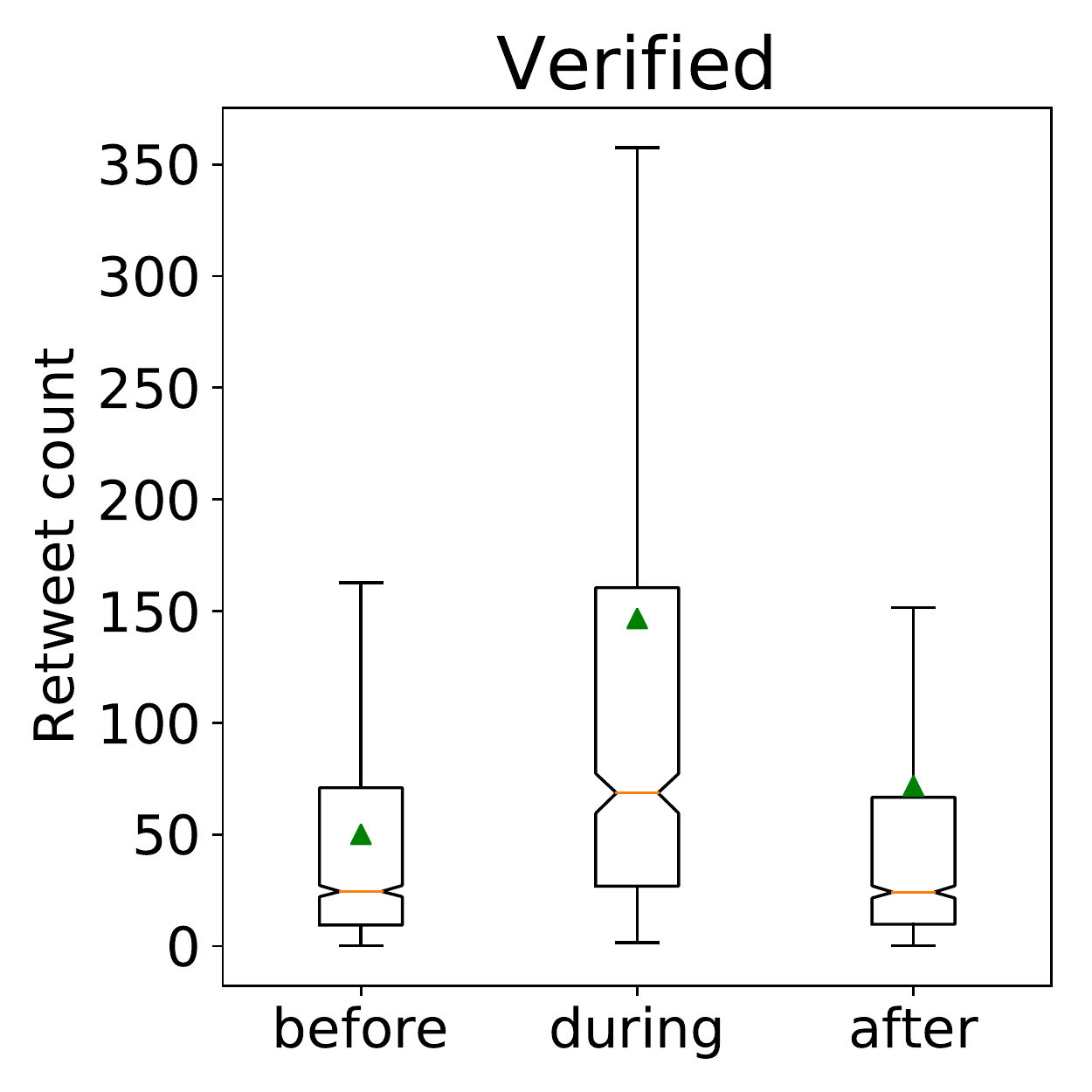}}
\end{minipage}%
\begin{minipage}{.24\linewidth}
\centering
\subfloat[]{\label{}
\includegraphics[width=\textwidth, height=\textwidth]{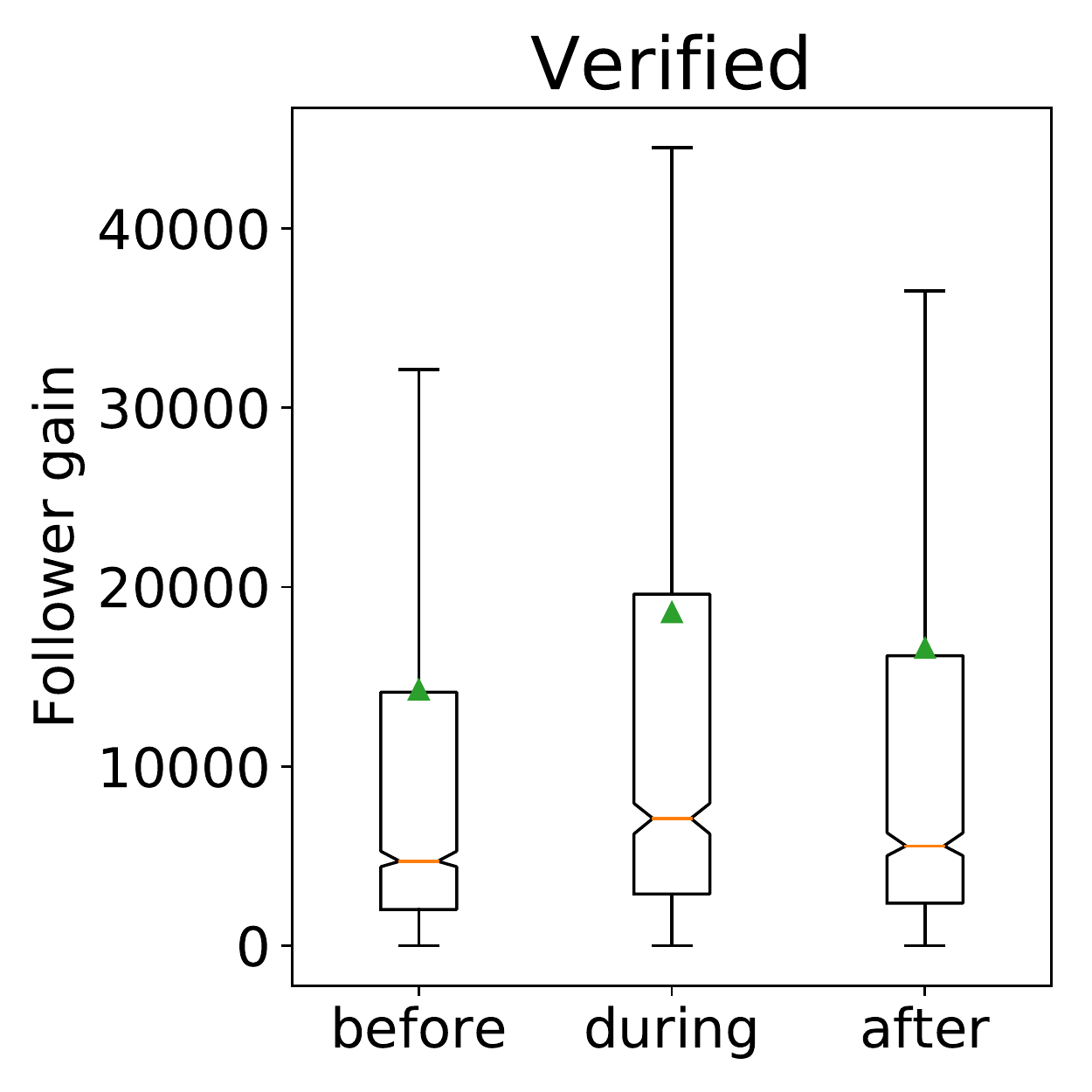}}
\end{minipage}%
\begin{minipage}{.24\linewidth}
\centering
\subfloat[]{\label{}
\includegraphics[width=\textwidth, height=\textwidth]{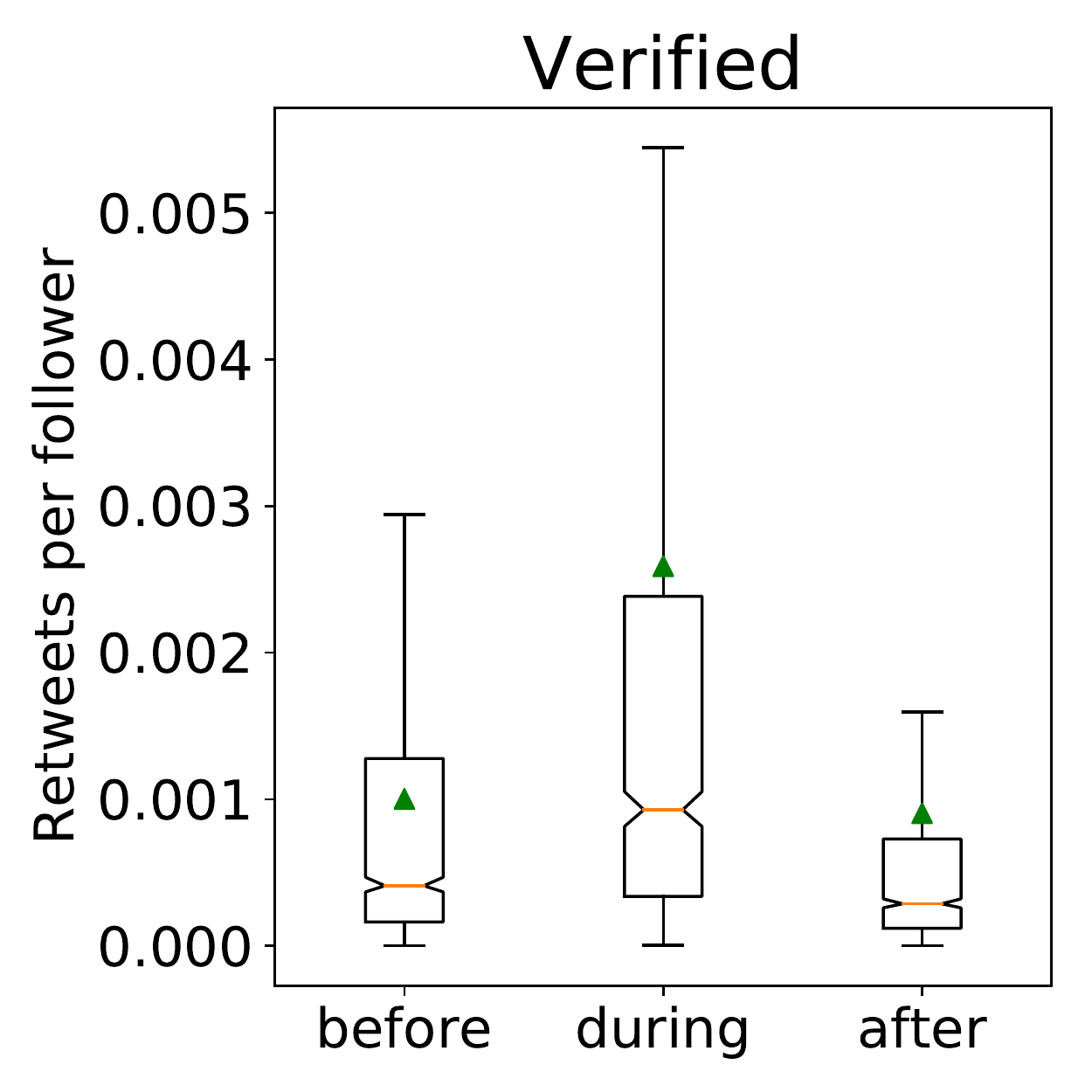}}
\end{minipage}%
\begin{minipage}{.24\linewidth}
\centering
\subfloat[]{\label{}
\includegraphics[width=\textwidth, height=\textwidth]{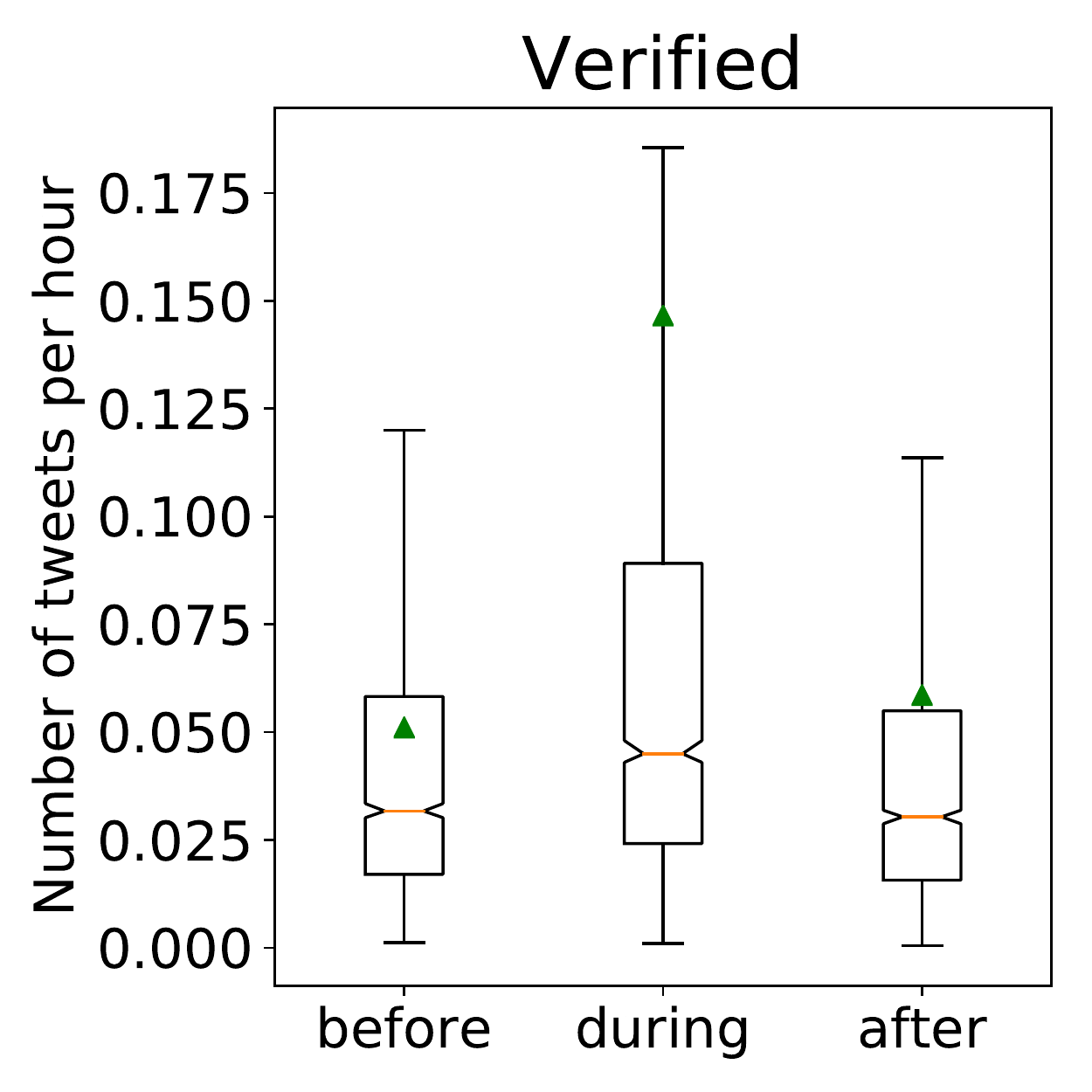}}
\end{minipage}%
\caption{Notched box plots showing the distributions before, during, and after a hot streak. The orange line shows the median, green triangle the mean. 
The notches indicate 95\% confidence intervals around the median obtained via bootstrapping. (a)~Retweet count, (b)~follower gain, (c)~retweet count per follower, (d)~number of tweets per hour.}
\label{fig:hotstreaks_foll_gain}
\end{figure*}

\spara{Retweet count} In Figure~\ref{fig:hotstreaks_foll_gain}(a), we see that the mean retweet count increases significantly\footnote{Significance computed using Welch's $t$-test.} 
during a hot streak when compared to the period before and after. This is expected by the definition of hot streaks. However, we also observe that the retweet counts after the hot streak are significantly higher than before, indicating some spillover effects.
    
\spara{Follower gain} We measure the growth in number of followers before, during, and after a hot streak.
Figure~\ref{fig:hotstreaks_foll_gain}(b) shows the follower gain.
We observe that the follower gain during a hot streak is significantly higher than before or after the hot streak.
This might be because more retweets lead to more followers~\cite{myers2014bursty}.
Similar to above, we also observe that the follower gain after a hot streak is significantly higher than the follower gain before.
    
\spara{Retweets per follower} To understand if the increased retweet count after the hot streak is due to the increased number of followers, we  look at the retweets per follower by dividing the average number of retweets before, during and after by the average number of followers during the same period. 
We can see from Figure~\ref{fig:hotstreaks_foll_gain}(c) that the increase after the hot streak vanishes in this case. Retweets per follower before and after the hot streak are significantly less frequent than during the hot streak, but there is no significant difference between before and after.

\spara{Activity} We look at the activity (all the tweets by the user) before, during, and after a hot streak. Note that this also includes retweets (unlike in Section~\ref{sec:identifying}), because we want to know how a user's overall tweeting behavior is associated with the impact created by their own tweets. To measure this, we compute the number of tweets per hour during, before and after a hot streak. We find that, during hot streaks, the activity is higher (Figure~\ref{fig:hotstreaks_foll_gain}(d)).
%


\subsection{Factors affecting hot streaks}
\label{sec:factors}
In the previous section, we saw that hot streaks are characterized by a higher retweet count, an increased follower gain, and an increase in activity.
In this section, we try to dig deeper into each of those factors and try  to understand the role they play in the hot streak phenomenon. 
We particularly look at three factors: network (behavior of followers and retweeters), content (tweets posted during a hot streak), and activity (number and type of tweets).
We want to measure how much of the hot streak characteristics we observed in the previous section are accounted for by these three factors.


\spara{Network}
In the previous section, we found that during a hot streak, users gain many followers, indicating changes to the network. To further analyze the role of the network during a hot streak, we look at individual retweeters who retweeted a user before, during, and after the hot streak.

\begin{figure}[ht]
\centering
\begin{minipage}{.49\linewidth}
\centering
\includegraphics[width=\textwidth, height=\textwidth]{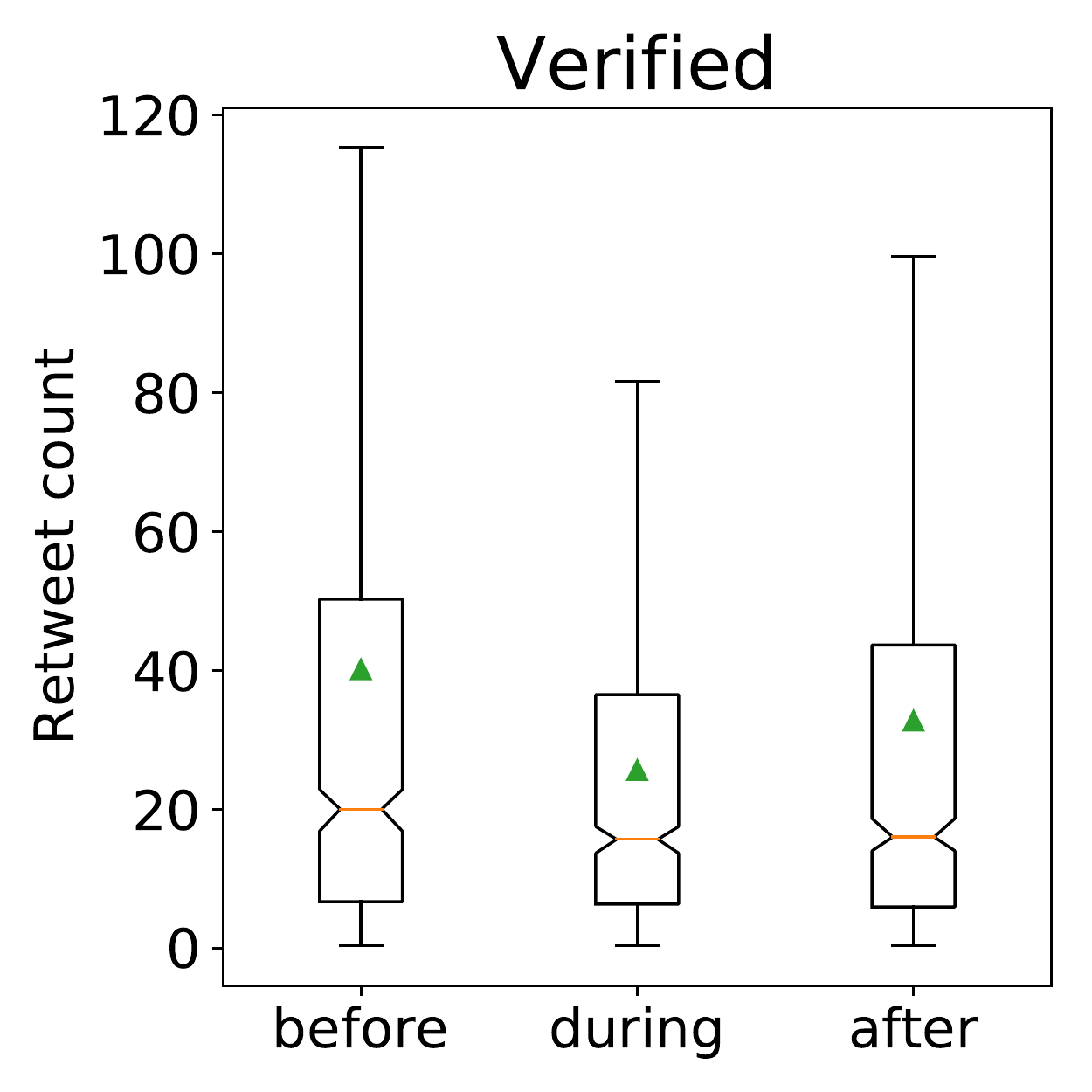}
\end{minipage}%
\begin{minipage}{.49\linewidth}
\centering
\includegraphics[width=\textwidth, height=\textwidth]{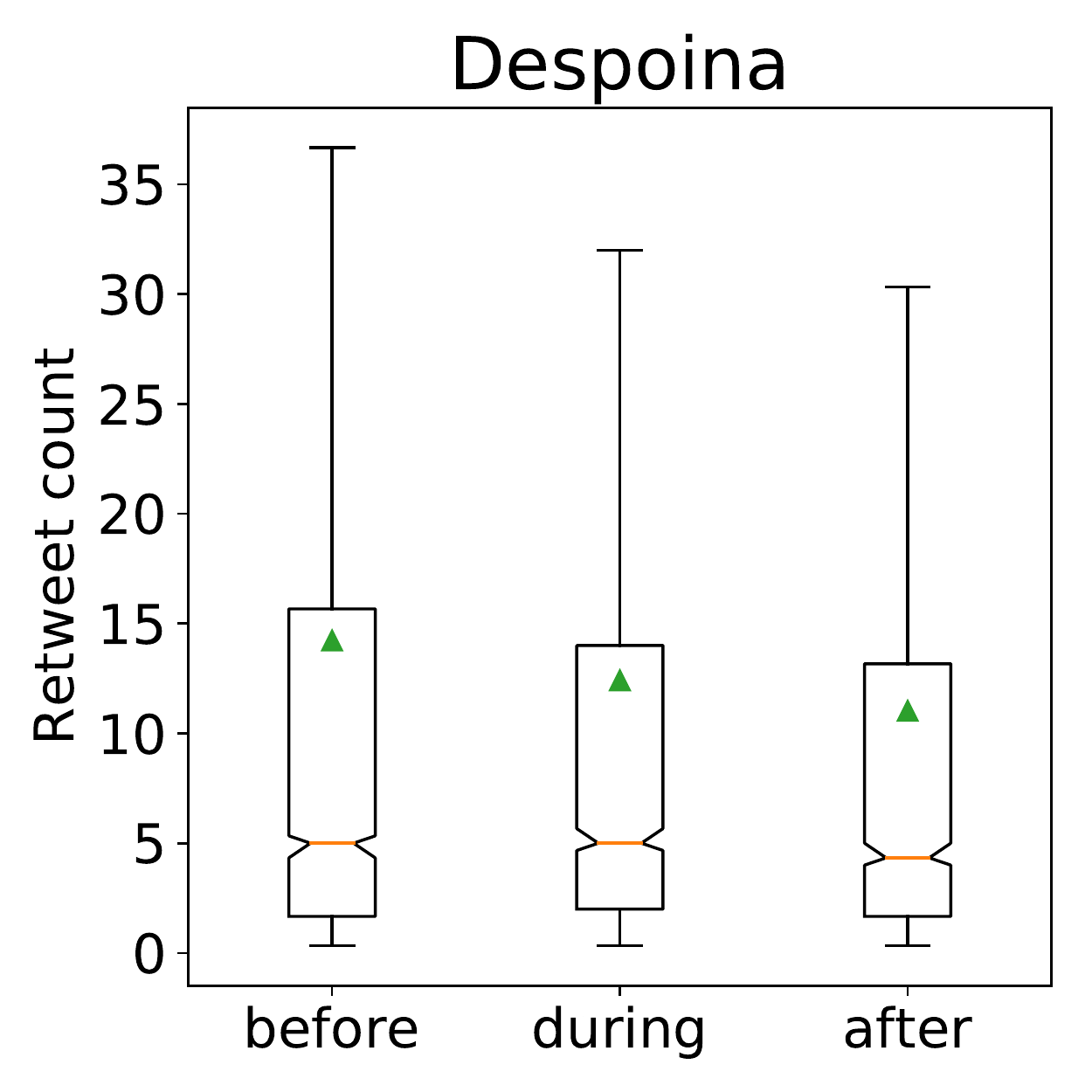}
\end{minipage}%
\caption{Retweet counts when only considering retweeters who retweeted before a hot streak.}
\label{fig:retweeters}
\end{figure}
\begin{figure}[t]
\centering
\begin{minipage}{.49\linewidth}
\centering
\includegraphics[width=\textwidth, height=\textwidth]{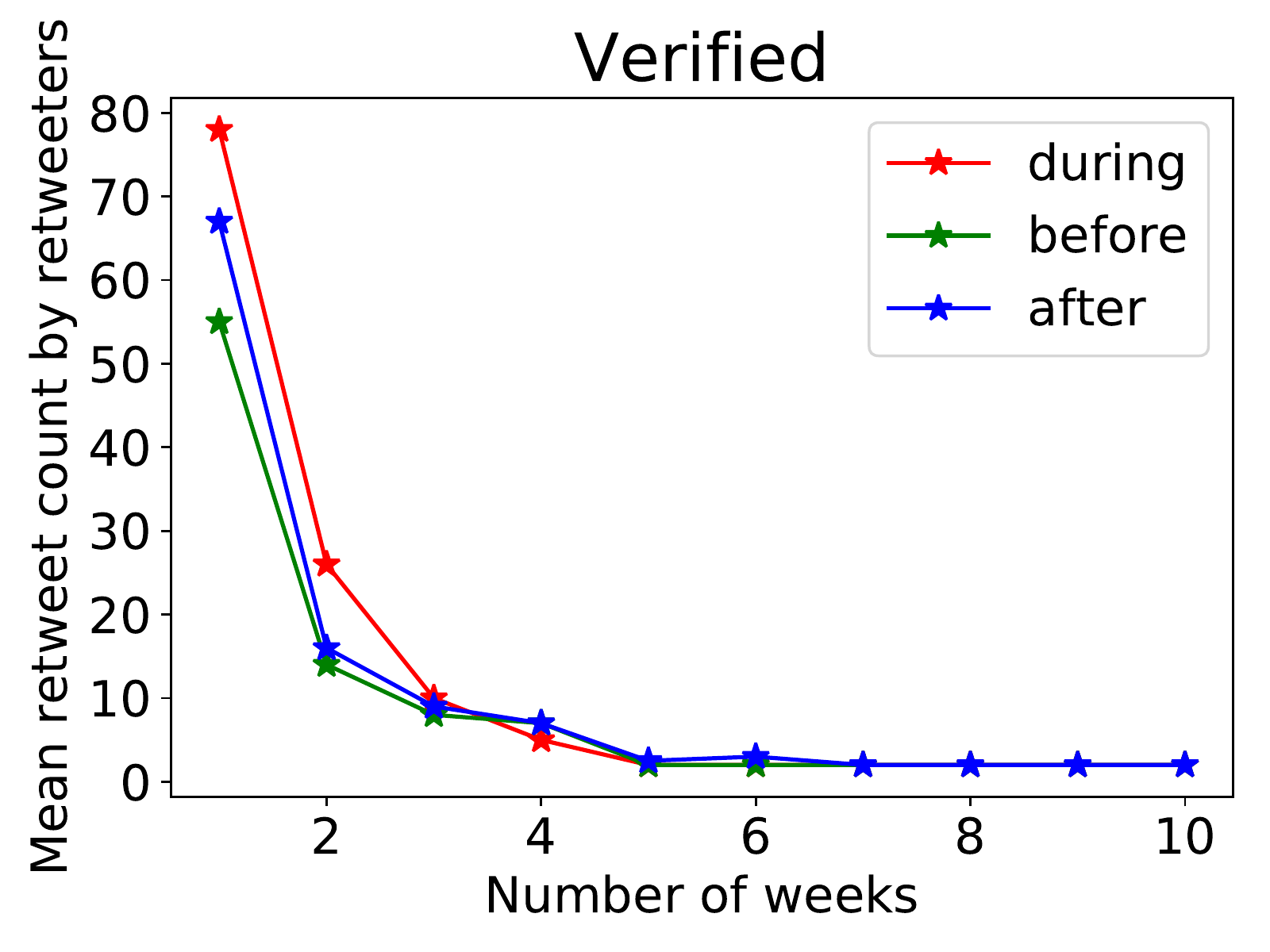}
\end{minipage}%
\begin{minipage}{.49\linewidth}
\centering
\includegraphics[width=\textwidth, height=\textwidth]{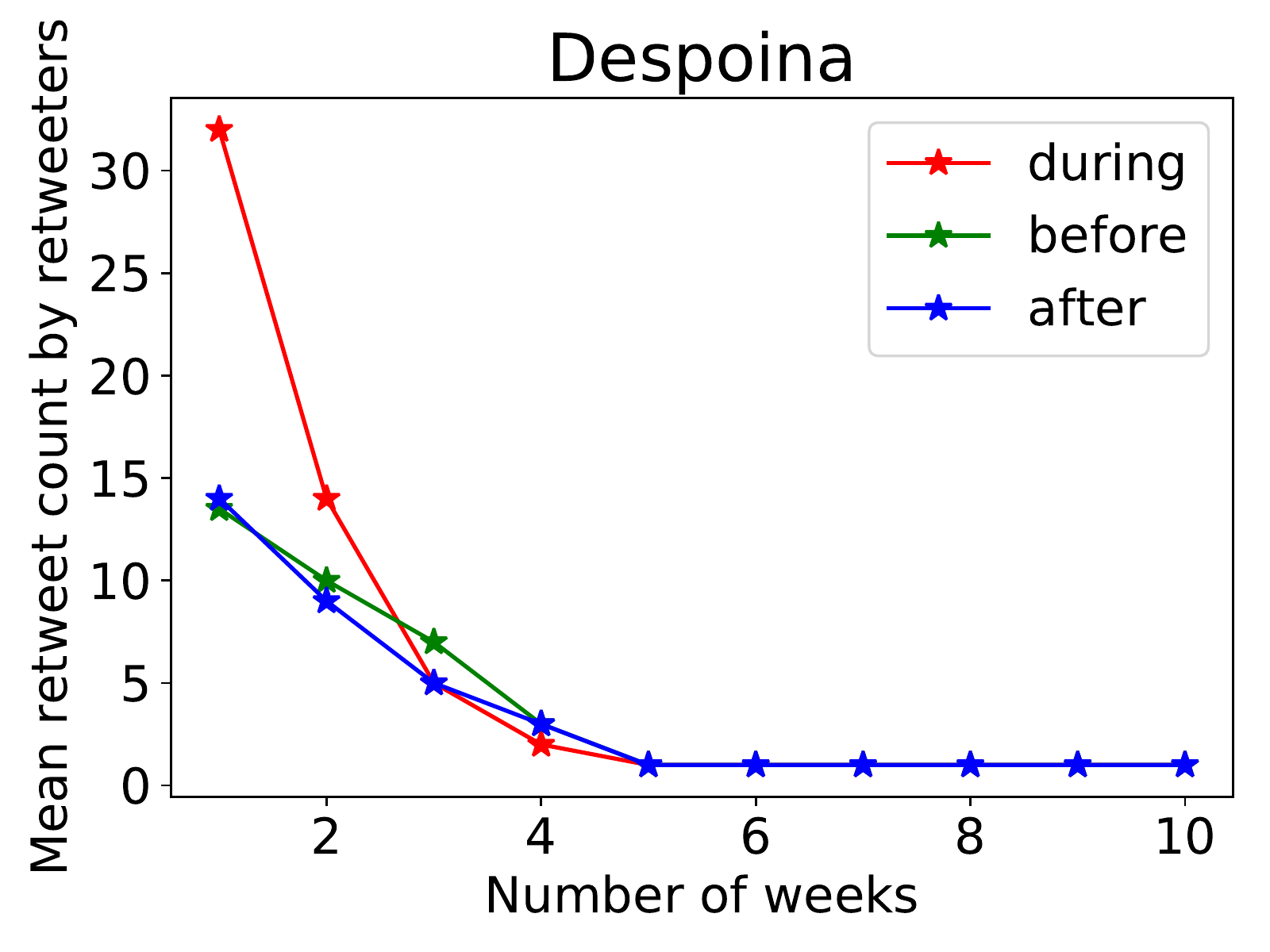}
\end{minipage}%
\caption{Retweet counts by retweeters starting before, during, and after
hot streaks. The $x$-axis shows the number of weeks since the beginning of the period before, during, or after the respective hot streak.
Retweeters starting during a hot streak have a higher retweet count early on, but it drops off quickly.}
\label{fig:retweeters_distribution}
\end{figure}

We fix the set of retweeters who retweeted the user before a hot streak and tracked their activity during and after the hot streak. Surprisingly, we find that these retweeters retweet the user significantly \textit{less} during a hot streak, compared to before. Figure~\ref{fig:retweeters} shows a box plot of retweeter behavior.
We also analyze the retweet count by only retweeters who started retweeting
during
the hot streak and compared that to retweeters who started
before or after the hot streak. We find that retweeters who start during the hot streak have a significantly higher retweet count (Figure~\ref{fig:retweeters_distribution}).
We can also see from Figure~\ref{fig:retweeters_distribution} that the retweeters who started retweeting
during
a hot streak 
lose interest more quickly than those who started retweeting before
or after.
These findings indicate that hot streaks are driven mostly by new retweeters who start retweeting the user suddenly, but lose interest quickly.


\spara{Content and activity} 
Above, we looked at \textit{who} is driving the increase in retweet count during a hot streak. 
Next, we dig deeper into content\hyp related factors, looking at \textit{what} drives such increase in impact.
To understand this, we looked at various aspects in terms of content posted and types of activities by users during a hot streak and compared it with periods before and after the hot streak.
Specifically, we looked at attributes from tweet patterns to measure activity features such as these:
fraction of tweets that are retweets\slash replies\slash mentions;
fraction of tweets containing media\slash URLs\slash hashtags;
tweet length;
and content related features such as entropy of the tweet in terms of the word distribution, entropy of the topic distribution, and average sentiment of tweets.
Topics were extracted using latent Dirchlet allocation (LDA)
as implemented in Python's scikit-learn package.
Sentiment was extracted using SentiStrength~\cite{thelwall2017heart}.
%

Table~\ref{tab:content_properties} shows the results comparing the period during the hot streak to periods of the same number of tweets before and after the hot streak.
Each column compares two values:
before \vs\ during,
during \vs\ after,
and before \vs\ after.
A cross (\xmark) indicates no statistically significant differences ($p < 0.01$). 
A ``\textgreater'' (``\textless'') indicates that the element on the left (right) in a column is significantly greater than the element on the right (left) for all the datasets.
For instance, the ``\textgreater'' in Table~\ref{tab:content_properties}, the first row, second column (``retweet, during \vs\ after'') indicates that the fraction of tweets that are retweets is significantly higher during the hot streak when compared to the period after the hot streak. 
Significance is tested using a $t$-test of means, and a Bonferroni correction is applied to the $p$-values. 
A value is marked ``\textgreater'' or ``\textless'' only if it holds across all four datasets.
We can observe the following results from Table~\ref{tab:content_properties}. During a hot streak, users have
(i)~a higher fraction of tweets that are retweets or contain media, 
(ii)~a lower fraction of tweets that are replies/mentions,
(iii)~longer tweets, with higher entropy both in terms of words and topics,
and
(iv)~no change in sentiment. 


\begin{table}[]
\caption{Content and activity features before, during, and after a hot streak. Rows: fraction of tweets that were retweets\slash replies\slash mentions; fraction of containing hashtags\slash URLs\slash media; tweet length; entropy of word distribution, entropy of topic distribution; average sentiment.
Each column compares pairs of values before/during/after a hot streak.
A cross (\xmark) indicates no statistically significant differences ($p < 0.01$). %
A ``\textgreater'' (``\textless'') indicates that the element on the left (right) in a column is significantly greater than the element on the right (left) for all the datasets.}
\label{tab:content_properties}
{\resizebox{\linewidth}{!}{
\begin{tabular}{c|c|c|c}
\hline
            & before \vs during    & during \vs after         & before \vs after       \\
\hline
retweet     &     \textless    &     \textgreater     &    \xmark        \\
reply       &    \textgreater       &     \textless    &   \xmark       \\
mention     &   \textgreater      &     \textless      &   \xmark        \\
hashtag     &   \xmark      &     \xmark      &   \xmark        \\
URLs         &       \xmark      &     \textless      &   \textless    \\
media       &       \textless      &     \textgreater      &   \xmark    \\
tweet length   &   \textless      &     \textgreater      &   \xmark        \\
entropy tweets     &     \textless      &     \textgreater      &   \xmark      \\
entropy topics     &     \textless      &     \textgreater    &   \xmark      \\
sentiment     &     \xmark      &     \xmark      &   \xmark      \\
\hline
\end{tabular}}}
\end{table}

\spara{Content and activity \vs network}
We saw that content, activity, and network features become significantly different during a hot streak. To understand the role of each of these factors, we look for users for whom there was no significant change in the network, i.e., users for whom there was no significant increase in follower count during a hot streak, and investigate their content and activity.
%
%
The following trends from Table~\ref{tab:content_properties} still persist  for this subset of users: (i)~lower fraction of replies/mentions, (ii)~higher fraction of media, and (iii)~higher entropy of topics.
The effects we find here show that irrespective of changes in the network, users exhibit a strong change in how they act and how they produce content during a hot streak. 

%
%
%

\spara{Predicting hot streaks}
We now build a simple binary classifier to predict for a user whether a week belongs to a hot streak or not, based on the three sets of features described above: content, activity, and network.
Concretely, we use the following features:
(a)~content and activity features: fraction of tweets that are retweets\slash replies\slash mentions, fraction of tweets containing hashtags\slash URLs\slash media, tweet length, entropy of the word distribution, entropy of the topic distribution (obtained using LDA), sentiment of the tweets;
(b)~network features: number of followers gained, average activity of the retweeters, number of retweeters tweeting for the first time, number of existing retweeters active during that week. 

Table~\ref{tab:prediction_hotstreaks} shows the accuracy of a random forest classifier for different sets of features on a class\hyp balanced dataset. The numbers in brackets show standard deviation over a 10-fold cross\hyp validation.
We also evaluate the classifier using just content/activity and just network features.
We see that both network and content/activity features individually do better than a random prediction (accuracy of 50\%). However, their combination improves the accuracy significantly. 
We computed the features that play a role in the prediction using the decrease in impurity in the random forest classifier.
Entropy of topics, number of followers, and number of tweets come up as the most important features in all the datasets.
It is interesting to see that the most important features span all three categories of features---content, network, and activity.
Our main goal with the prediction task was to determine which factors are most associated with hot streaks,
by looking at the features that would best predict hot streaks.
However, the answer is not clear. A mix of all the three feature sets is involved, along with some external noise which we cannot account for (constituting the remaining 30\% in accuracy in our model), pointing to the limits of predicting real-world phenomena~\cite{martin2016exploring}.

%




\begin{table}[]
\caption{Accuracy predicting whether a particular week is in a hot streak or not. ``Content'' and ``Network'' refer to just content\slash activity- and network\hyp related features in isolation. ``Combined'' indicates all features combined.}
\label{tab:prediction_hotstreaks}
\begin{tabular}{l|l|l|l}
\hline
                                 & Content     & Network     & Combined    \\
\hline
\verified         & 0.67 (0.03) & 0.62 (0.02) & 0.72 (0.04) \\
\political        & 0.72 (0.01) & 0.63 (0.01) & 0.78 (0.01) \\
\despoina         & 0.72 (0.0)  & 0.63 (0.0)  & 0.77 (0.0)  \\
\begin{tabular}[c]{@{}l@{}}\cpt{Despoina}\\ (\cpt{random})\end{tabular} & 0.70 (0.02) & 0.63 (0.01) & 0.75 (0.01)
\end{tabular}
\end{table}



\section{Impact around the most retweeted tweet}
\label{sec:before_after}

In the previous section, we looked at trends in impact over the entire career of a user. In this section, we zoom into the career and understand the dynamics of retweet activity surrounding the \textit{most impactful} tweets of a user.

\begin{figure}[t]
\centering
\begin{minipage}{.49\linewidth}
\centering
\includegraphics[width=\textwidth, height=\textwidth]{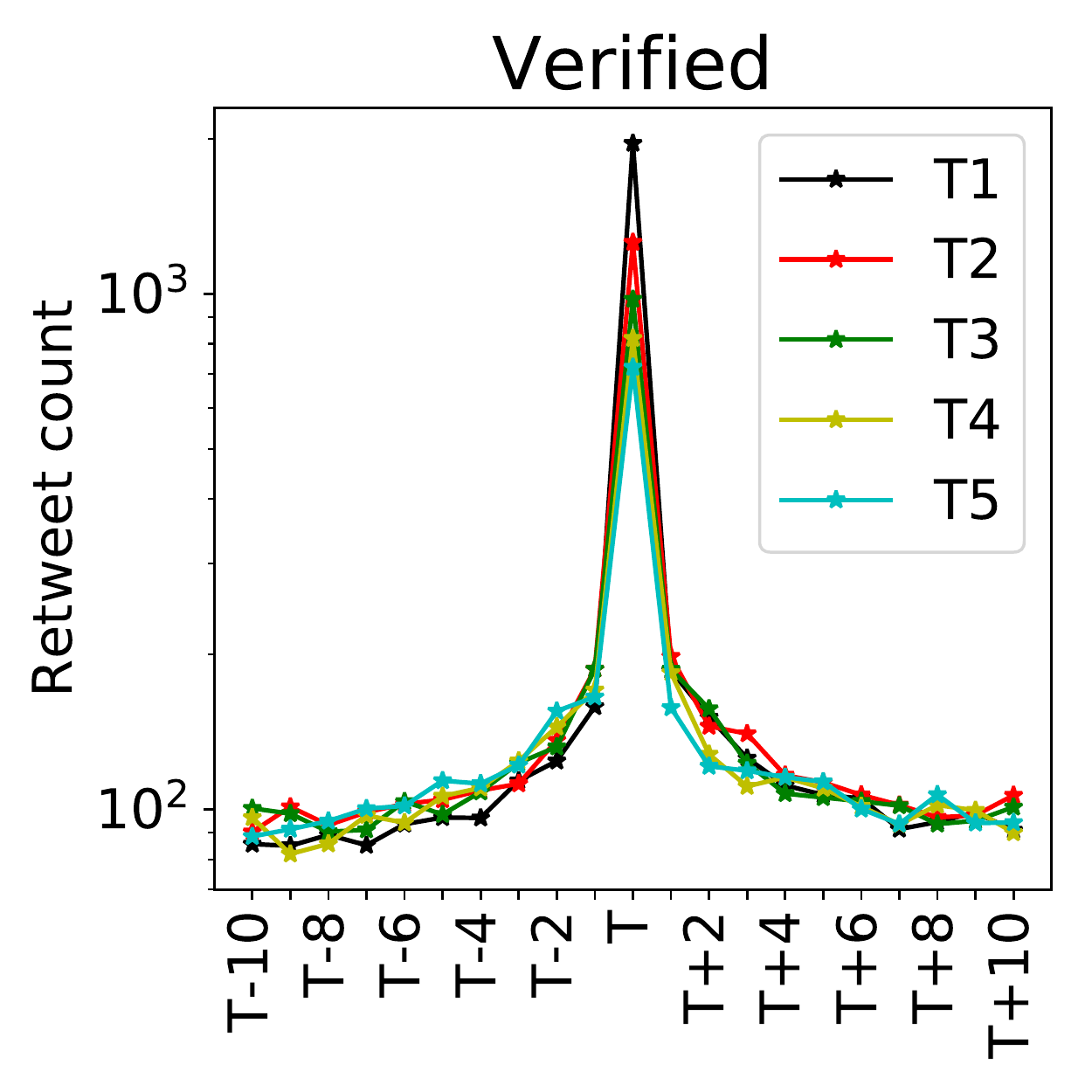}
\end{minipage}%
\begin{minipage}{.49\linewidth}
\centering
\includegraphics[width=\textwidth, height=\textwidth]{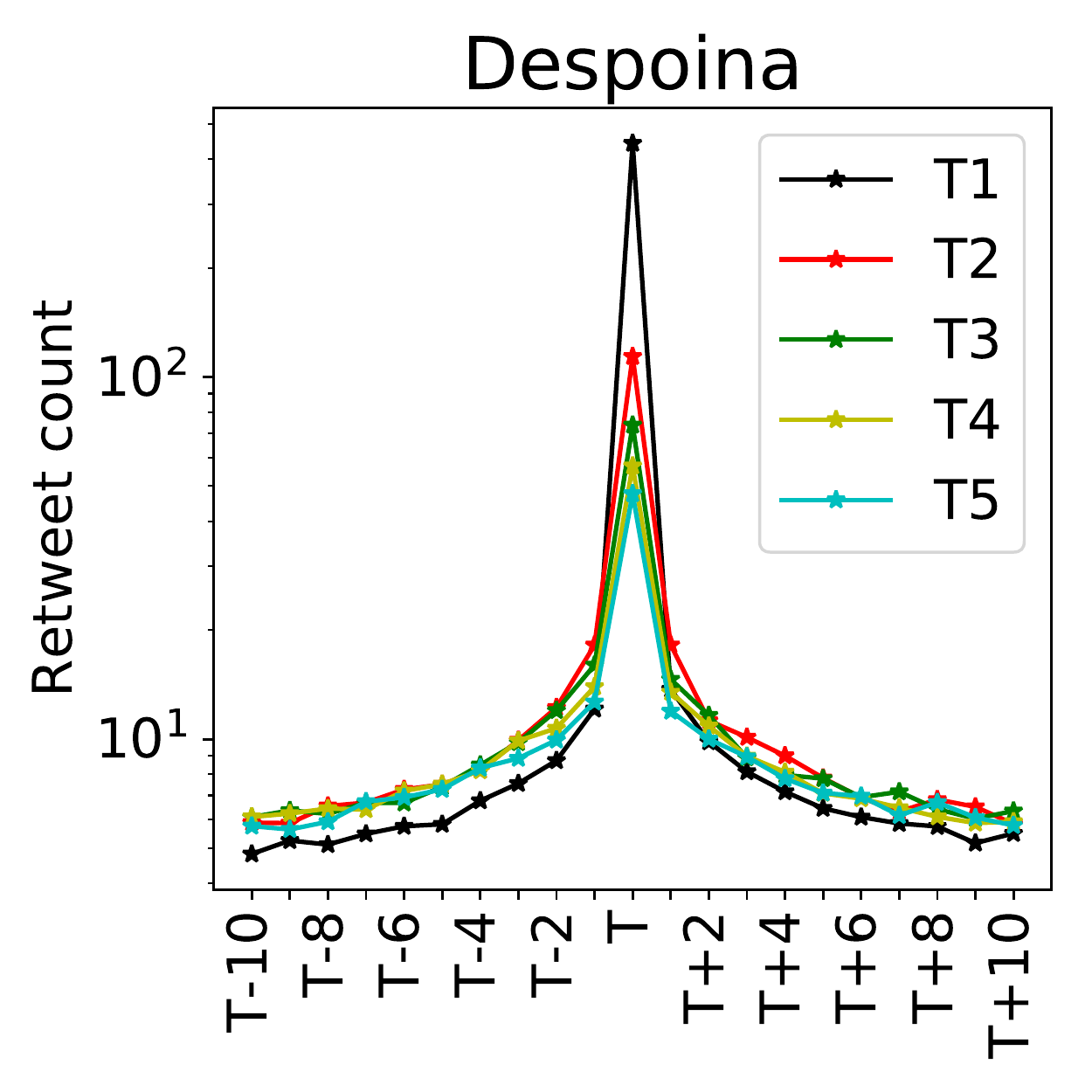}
\end{minipage}%
\par\medskip
\begin{minipage}{.49\linewidth}
\centering
\includegraphics[width=\textwidth, height=\textwidth]{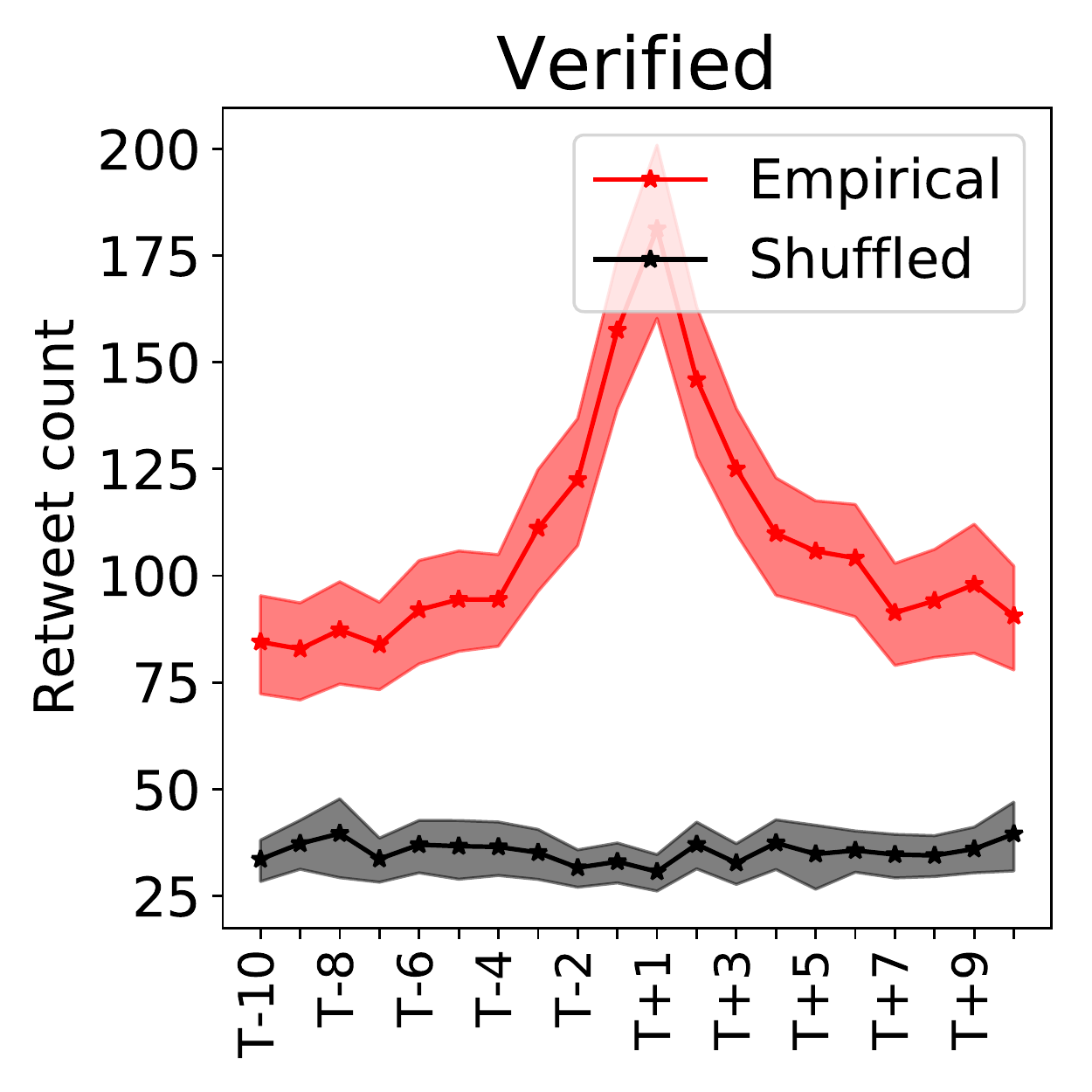}
\end{minipage}%
\begin{minipage}{.49\linewidth}
\centering
\includegraphics[width=\textwidth, height=\textwidth]{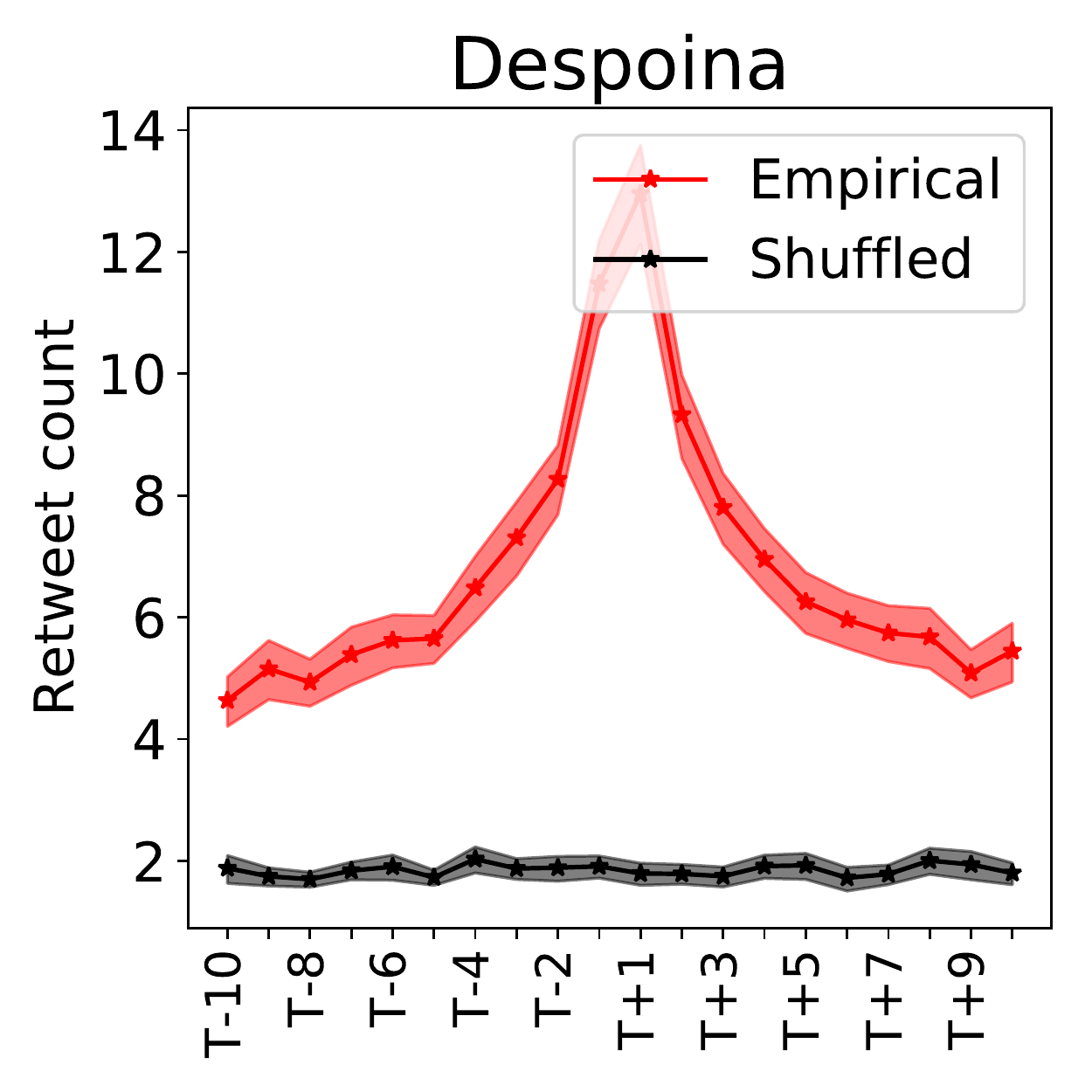}
\end{minipage}%
\caption{Top row: Mean retweet counts of 10 tweets before ($T-10,\dots,T-1$) and after ($T+1,\dots,T+10$) each of the top 5 most retweeted tweets ($T$) per user. Bottom row: The 10 tweets before and after, without the focal tweet ($T$) itself, compared to a shuffled timeline (in black), with 95\% confidence intervals.}
\label{fig:hotstreaks_before_after}
\end{figure}

Looking at the period of 10 tweets before and after the most retweeted tweet of a user, we found that the most retweeted tweet does not occur in isolation, but that there is a build-up to it.
The top row in Figure~\ref{fig:hotstreaks_before_after} shows the mean retweet count across all users for the five most retweeted tweets ($T_1, \dots, T_5$) and the 10 tweets preceding/succeeding them.
The bottom row just shows the 10 tweets before and after the most retweeted tweet ($T_1$), excluding $T_1$.
We can clearly see that there is a build-up towards, and drop-off after, the most retweeted tweet.
This pattern holds for various stratifications of the users, along experience, retweet counts, and datasets.
Comparing this pattern with 10 tweets before and after the most retweeted tweet in a shuffled career, we see from Figure~\ref{fig:hotstreaks_before_after} (bottom row, black line) that  the trend disappears.

%

\begin{figure}[ht]
\centering
\begin{minipage}{.49\linewidth}
\centering
\includegraphics[width=\textwidth, height=\textwidth]{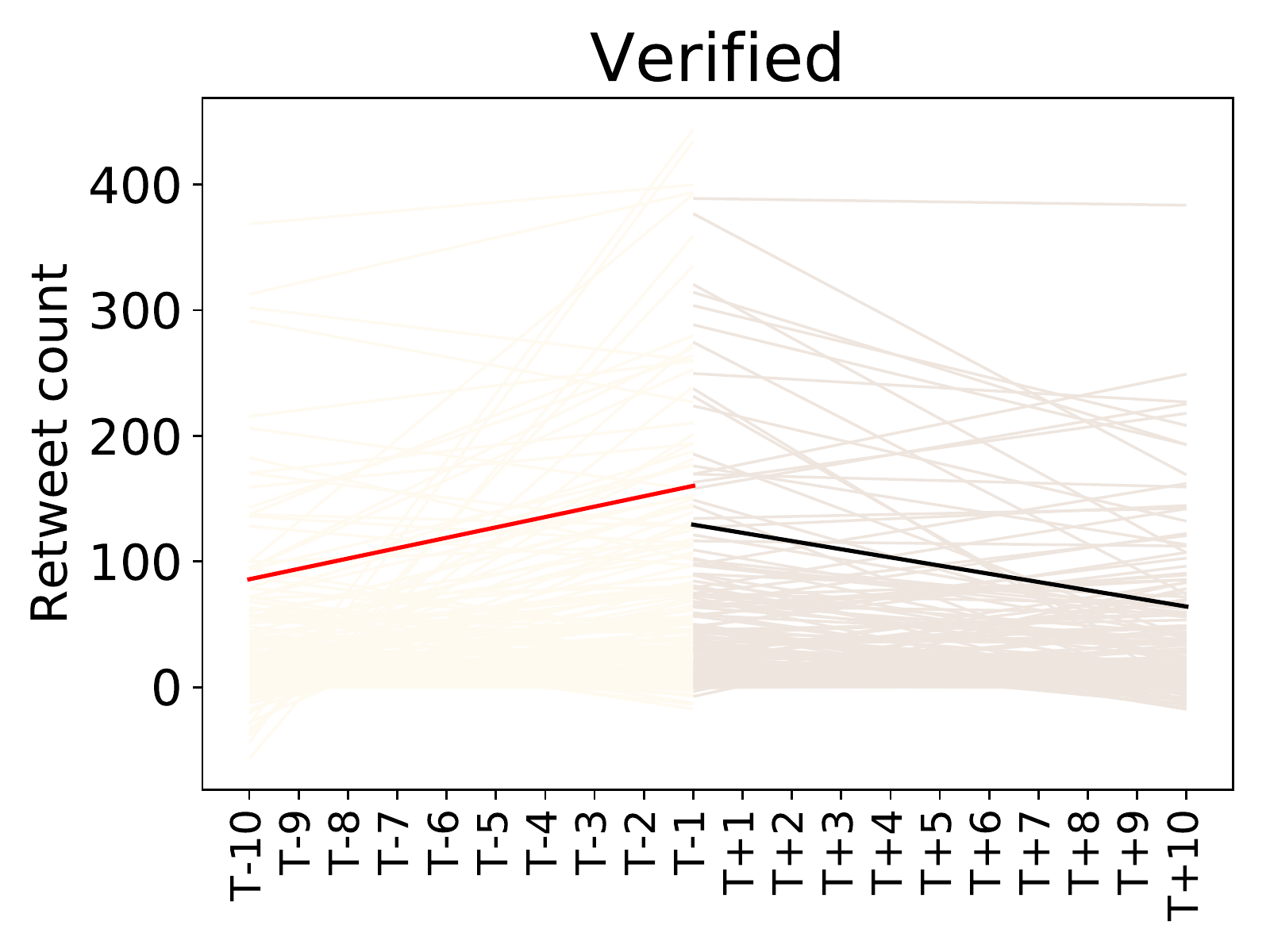}
\end{minipage}%
\begin{minipage}{.49\linewidth}
\centering
\includegraphics[width=\textwidth, height=\textwidth]{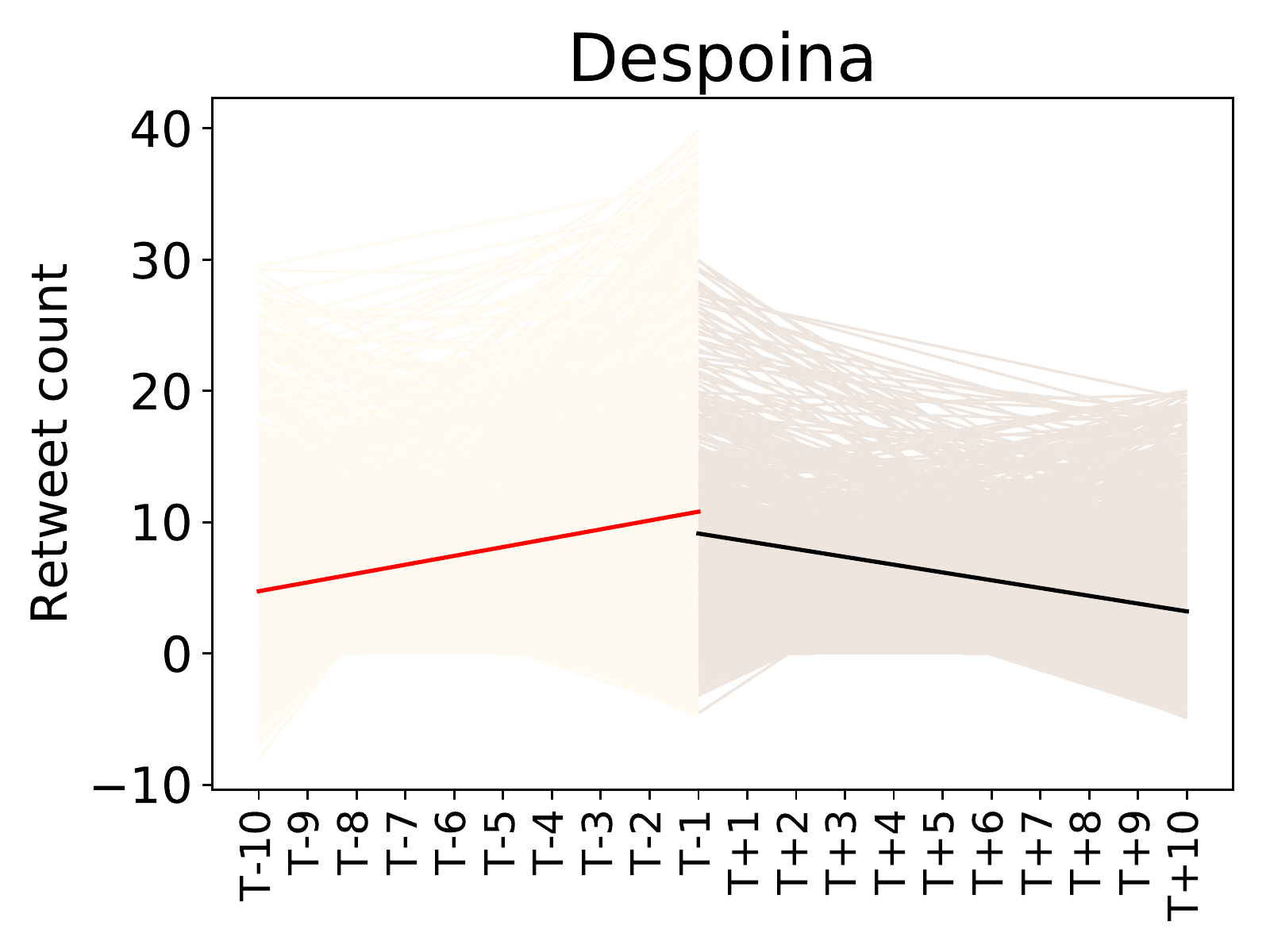}
\end{minipage}%
\caption{Linear fits for retweet counts before and after $T_1$ (the user's most retweeted tweet), for two datasets. We see that the average slope (dark) is positive (negative) before (after) $T_1$, indicating a clear build-up (drop-off) around $T_1$.}
\label{fig:slopes}
\end{figure}

To understand if this is a prevalent phenomenon across users, 
we perform linear fits for the retweet counts of 10 tweets before and after the most retweeted tweet $T_1$ (excluding $T_1$ itself) and plot the average slope of these lines.
Figure~\ref{fig:slopes} shows the average slope of the increase (in red) and decrease (in black). We also show the individual fits for all users (in light shades). 
We can see from Figure~\ref{fig:slopes} that for most users (over 60\%) there is an average positive (negative) trend before (after) $T_1$.

\begin{figure}[ht]
\centering
\begin{minipage}{.49\linewidth}
\centering
\includegraphics[width=\textwidth, height=0.625\textwidth]{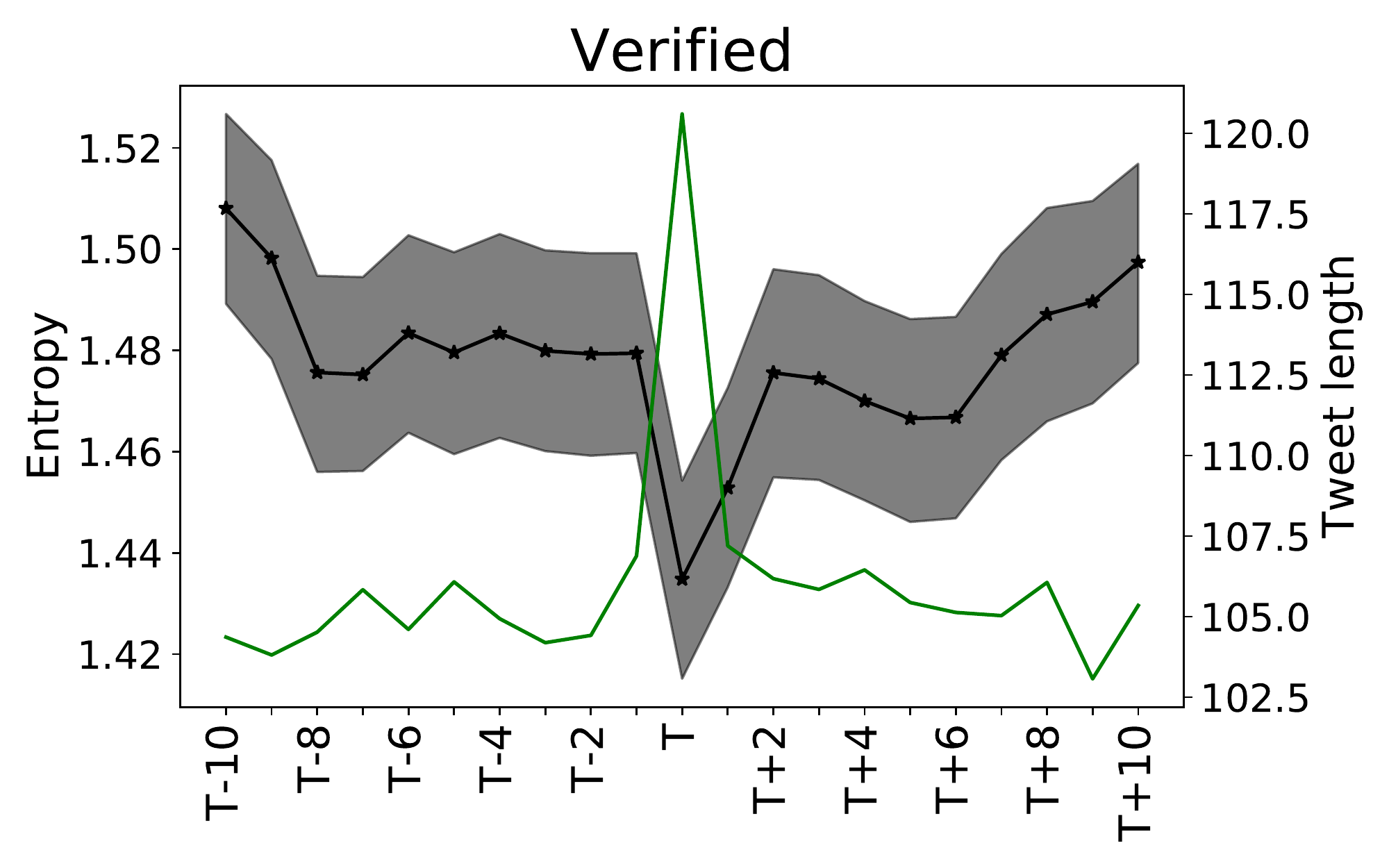}
\end{minipage}%
\begin{minipage}{.49\linewidth}
\centering
\includegraphics[width=\textwidth, height=0.625\textwidth]{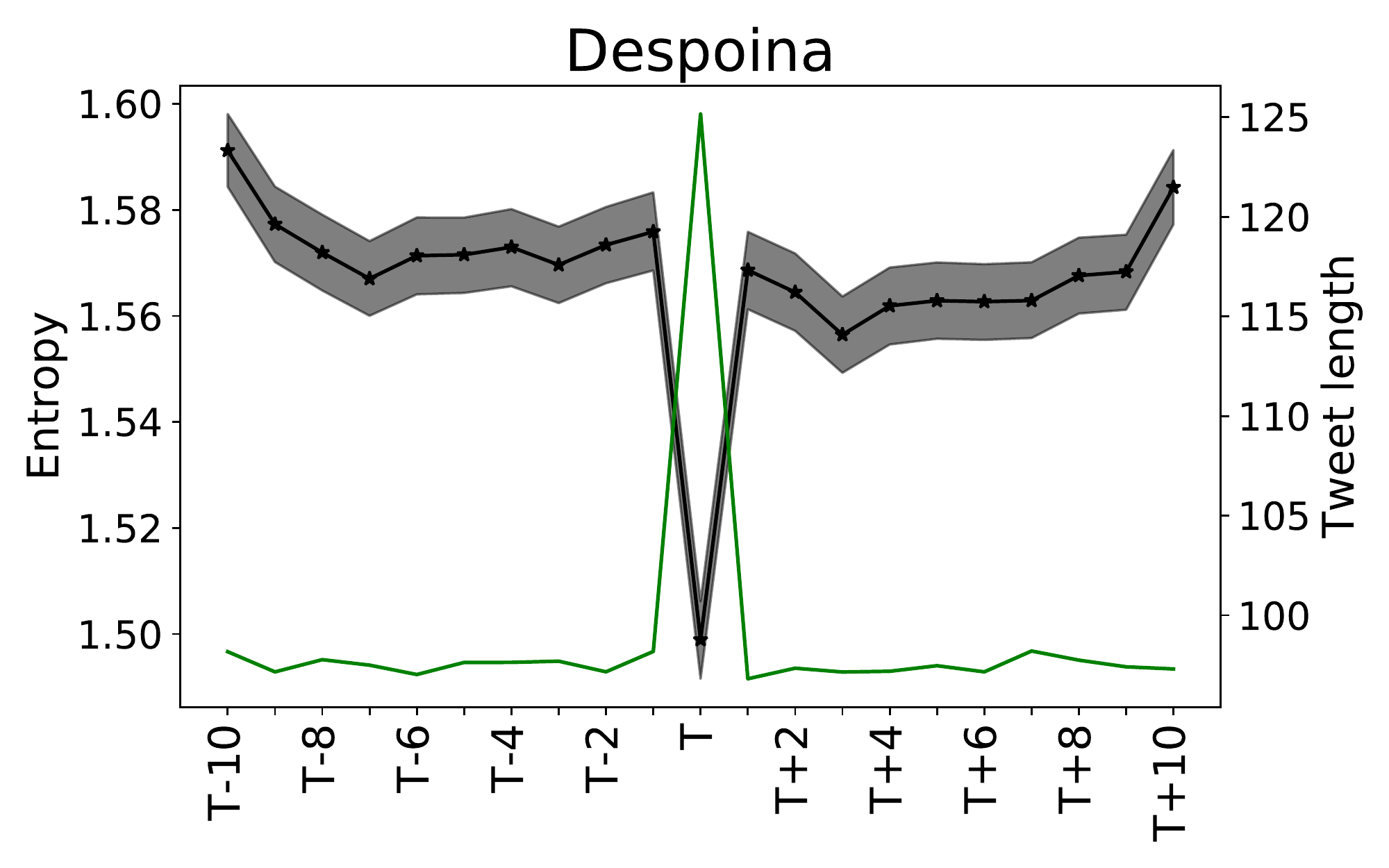}
\end{minipage}%
\caption{Entropy of the topic distributions before, during, and after $T_1$ (the user's most retweeted tweet) along with 95\% confidence intervals (black line). The green line shows the average length of the tweet (on the $y$-axis on the right).}
\label{fig:entropy_before_after}
\end{figure}

We further look at the role of content in this phenomenon, by
computing the entropy of the topic distribution of the 10 tweets before/after $T_1$. 
Figure~\ref{fig:entropy_before_after} shows that  the entropy of the topic distribution decreases specifically at $T_1$.  
To understand if this is just an artifact of tweets being shorter, we also plotted the average length of the tweets before and after $T_1$.
We find that even though the tweets are longer at $T_1$, the entropy of topics is lower.
We looked at whether these tweets (10 before/after) are part of a thread on Twitter, but could not find any evidence for that.
Future research should analyze the content of most retweeted tweets more closely in order to explain the marked decrease in entropy.

\section{Related work}
\label{sec:related}

\spara{Hot hand} 
The ``hot hand'' hypothesis originates from basketball, where a shooter is allegedly more likely to score if their previous attempts were successful,  while having ``hot hands''. 
\textcite{gilovich1985hot} were the first to show that there is no evidence of hot hands in basketball, arguing that short random sequences occur by chance and are not representative of their generating process.
However, these claims have been refuted in numerous studies later, pointing to flaws in the statistical analysis in the original study and showing the existence of hot hands on different datasets~\cite{miller2018surprised}.
Similar phenomena exist in other domains, too, such as gambling~\cite{ayton2004hot}.
See the supplementary material of \textcite{liu2018hot} for a thorough review of studies on hot hands and the proposed potential reasons for this phenomenon.
\commentAA{After reading the related work section, I am not convinced about the reasonable differentiation of this work from say the state-of-the-art for hot-streaks say Liu et al. 2018. It clearly feels that the idea has been borrowed, although you state that there are differences, but that part does not read that significant as of now. I think it should be made more clear, and also an additional set of challenges should be added, that why doing such a study on social media is much much harder, and not merely an application of ideas from previous work.}
Perhaps most relevant to our work is recent research by \textcite{liu2018hot}, who show that in many disciplines such as science, art, and movies, careers exhibit hot streaks of increased creative and impactful output. 

Our work is different from the above studies in many ways. Firstly, we study entire careers of social media users looking at impact, rather than activity.
Our contribution is a combination of identifying trends in impact of a user in her career and showing for the first time the presence of the hot hand phenomenon on social media.
Secondly, an important aspect that is missing from all the above analysis is social influence, where the performance/impact of a person depends not only on her personal abilities, but also on who is judging them.
Our analysis on social media impact takes that into account. 
We try to untangle the relationship between content, activity, and network and show that each of these factors plays their own role in hot streaks.
Thirdly, social media is inherently different than domains such as art and science. 
For instance, activity on social media, unlike in science, sports, arts, or any other skill\hyp based occupation, has a much lower barrier. Being productive in terms of the number of tweets produced is easier than producing scientific papers or art work.
This might explain why some of our results are different from \textcite{liu2018hot}.

\spara{Bursts of activity on social media}
%
Sudden bursts of activity on Twitter have been studied previously. 
A sudden surge in the number of tweets/retweets were used to identify events~\cite{chierichetti2014event}.
\textcite{zhao2015seismic} looked at the distribution of such bursts for retweets from a tweet. They conclude that the interest in the tweet follows a power-law decay.
\textcite{myers2014bursty} show the impact of a burst of retweets on follower growth. They study how the follower network changes following a burst of retweet activity.
\textcite{barabasi2005origin} proposes that human activity in general is characterized by burstiness, with periods of high activity followed by rest periods. However, the grounds for such phenomena are not yet clearly understood. 
\textcite{gandica2016origin} show the existence of such burstiness in the behavior of editors on Wikipedia.
\textcite{goel2015structural} study virality on social media and show that modeling what goes viral and how is hard to predict due to the lack of proper structure in the nature of cascades.
%
%
%
%
%
Our work adds to all these fields. First, we show that there are bursty periods of impact in a user's career on social media, where, unlike Wikipedia~\cite{gandica2016origin}, the network factor plays an important role.
We also try to untangle content and network in this bursty behavior~\cite{myers2014bursty}.
%
Second, most of the above work analyzes single pieces of content (tweets) and the factors that influence their impact. In this paper, we go beyond a single tweet and look at the entire career of a user. \commentAA{something missing in this line: "the factors that influence impact a content's trajectory".}
%
%
The difference is that individual content can have different properties, like becoming viral or having a higher quality, but on average these effects might cancel out when looking at a lot of posts for a single user, hence giving rise to new patterns.
Finally, our results show that there is a pattern in the way individual impact is created. These findings could be useful in understanding collective behavior such as event detection, measuring influence, or virality.
%
%




\section{Discussion}
\label{sec:conclusions}

In this work, we analyze individual impact on Twitter using datasets consisting of careers of a large set of users sampled to represent a wide range of properties. 
We present new insights on the prevalence of structure and patterns in the impact in a user's career.
By showing that the results hold on multiple datasets and are robust to shuffling careers, we establish that these results are inherent properties of social media and not artifacts of sampling.

\spara{Clustering of impact}
We first show that the most impactful tweets of a user occur close together temporally for a large set of users. This result does not hold when user's careers are shuffled, indicating that real careers are characterized by such effects.
\commentAA{occur close together "in time" or "temporally".}

\spara{Hot streaks}
Next, we generalize the above finding and show that there is not just clustering of individual popular tweets, but extended periods of consistently high impact, which we call hot streaks.
Our analysis of the length of hot streaks shows that most users have hot streaks of around 10--20 tweets, and some users, of over 100 tweets.
We also show that retweets obtained during hot streaks do not constitute a major chunk of the total retweets accrued by a user.
Looking at how various factors affect a hot streak, we find significant differences in the network, content, and activity of a user.
%
%
%
%
A classifier to predict whether a week belongs to a hot streak or not performs reasonably well, with an accuracy close to 75\%.\commentAA{"affecting" -> "affect". Typo}

\spara{Impact around the most retweeted tweet}
Finally, we show that there is a build-up and drop-off surrounding a user's most retweeted tweets.
We tried to quantify if this is a prevalent phenomenon by looking at the slopes of linear fits for the 10 tweets before and after the most retweeted tweet, finding that this is a common trend among a majority of users.
%
Even though individual tweets may become viral at random~\cite{martin2016exploring}, such underlying structure around the most retweeted of an individual gives hope for predicting which tweets might become viral. 


\spara{Limitations}
This work has several limitations.
%
%
First, our datasets are not completely randomly sampled, and hence we cannot be sure about the generalizability of the results. Our datasets are picked in such a way to ensure at least some activity on the users' behalf. So the trends reported here may not hold for a random Twitter user.
However, our findings are novel and point towards structural properties in impact for an important subset of users.
%

\commentAA{"can not" -> "cannot"}
Second, our study
cannot fully explain the mechanisms by which
these effects occur.
%
The underlying process could be a complex mix of the features we considered along with other external, dataset\slash domain\slash time\hyp specific factors.
%
It may well be that some of this phenomena cannot be explained with observable features~\cite{martin2016exploring}, and trying to find explanations could be futile~\cite{langer1975illusion}.

\spara{Future work} 
The results presented in this paper are just one step towards understanding how user behavior evolves on social media. The large, rich, long-term dataset we created provides immense opportunities to explore many other aspects further, including downstream activities such as using these findings to study virality, influence, etc.
We looked at individual characteristics of impact. It would be interesting to see if such individual patterns propagate in the network. For instance, if a user is in a hot streak, do users in her network also experience hot streaks?
%
In preliminary work,
we observed similar phenomena on other social network platforms, such as YouTube and Instagram, though the effects were less pronounced and did not hold for a majority of the users.
Hence, future work should investigate how such trends generalize across platforms.


\bibliographystyle{aaai}
\bibliography{biblio}

\end{document}